\documentclass[longauth]{aa}

\usepackage{graphicx}
\usepackage{natbib}
\usepackage{scalerel}

\usepackage[table]{xcolor}

\usepackage{txfonts}

\usepackage[pdfencoding=auto,psdextra]{hyperref}
\hypersetup{
    colorlinks=true,
    linkcolor=blue,
    filecolor=magenta,
    urlcolor=blue,
    citecolor=blue
}
\makeatletter
\renewcommand*\aa@pageof{, page \thepage{} of \pageref*{LastPage}}
\makeatother

\usepackage[utf8]{inputenc}

\usepackage[switch, modulo]{lineno}

\renewcommand{\linenumbers}{}

\usepackage{euclid}

\defcitealias{Q1-Bamford}{B26}

\begin{document} 

\title{Euclid Quick Data Release (Q1)}
\subtitle{XLVII. The shape of cluster haloes revealed by intracluster light}

\newcommand{\orcid}[1]{} 
\author{Euclid Collaboration: T.~Kolcu\orcid{0000-0002-3365-9210}\thanks{\email{icl-group@nottingham.ac.uk}}\inst{\ref{aff1}}
\and N.~A.~Hatch\orcid{0000-0001-5600-0534}\inst{\ref{aff1}}
\and A.~Fernandez\orcid{0009-0001-4673-526X}\inst{\ref{aff1}}
\and S.~P.~Bamford\orcid{0000-0001-7821-7195}\inst{\ref{aff1}}
\and H.~Gao\orcid{0000-0003-1015-5367}\inst{\ref{aff1}}
\and J.~B.~Golden-Marx\orcid{0000-0002-6394-045X}\inst{\ref{aff1}}
\and G.~Martin\orcid{0000-0003-2939-8668}\inst{\ref{aff1}}
\and Y.~M.~Bah\'{e}\orcid{0000-0002-3196-5126}\inst{\ref{aff1},\ref{aff2}}
\and S.~L.~Ahad\orcid{0000-0001-6336-642X}\inst{\ref{aff3},\ref{aff4}}
\and C.~Bellhouse\orcid{0000-0002-6179-8007}\inst{\ref{aff1}}
\and M.~Butler\orcid{0009-0004-9273-4630}\inst{\ref{aff1}}
\and S.~de~Nicola\orcid{0000-0001-6173-5404}\inst{\ref{aff5},\ref{aff6}}
\and F.~Durret\orcid{0000-0002-6991-4578}\inst{\ref{aff7}}
\and A.~Ellien\orcid{0000-0002-1038-3370}\inst{\ref{aff8}}
\and A.~H.~Gonzalez\orcid{0000-0002-0933-8601}\inst{\ref{aff9}}
\and H.~Gully\orcid{0000-0001-7616-7278}\inst{\ref{aff1}}
\and M.~Kluge\orcid{0000-0002-9618-2552}\inst{\ref{aff5}}
\and M.~Maturi\orcid{0000-0002-3517-2422}\inst{\ref{aff10},\ref{aff11}}
\and M.~Montes\orcid{0000-0001-7847-0393}\inst{\ref{aff12}}
\and L.~Moscardini\orcid{0000-0002-3473-6716}\inst{\ref{aff13},\ref{aff14},\ref{aff15}}
\and R.~Ragusa\inst{\ref{aff16}}
\and A.~Saro\orcid{0000-0002-9288-862X}\inst{\ref{aff17},\ref{aff18},\ref{aff19},\ref{aff20},\ref{aff21}}
\and M.~Urbano\orcid{0000-0001-5640-0650}\inst{\ref{aff1},\ref{aff22}}
\and N.~Aghanim\orcid{0000-0002-6688-8992}\inst{\ref{aff23}}
\and B.~Altieri\orcid{0000-0003-3936-0284}\inst{\ref{aff24}}
\and S.~Andreon\orcid{0000-0002-2041-8784}\inst{\ref{aff25}}
\and N.~Auricchio\orcid{0000-0003-4444-8651}\inst{\ref{aff14}}
\and C.~Baccigalupi\orcid{0000-0002-8211-1630}\inst{\ref{aff18},\ref{aff19},\ref{aff20},\ref{aff26}}
\and M.~Baldi\orcid{0000-0003-4145-1943}\inst{\ref{aff27},\ref{aff14},\ref{aff15}}
\and S.~Bardelli\orcid{0000-0002-8900-0298}\inst{\ref{aff14}}
\and P.~Battaglia\orcid{0000-0002-7337-5909}\inst{\ref{aff14}}
\and R.~Bender\orcid{0000-0001-7179-0626}\inst{\ref{aff5},\ref{aff28}}
\and A.~Biviano\orcid{0000-0002-0857-0732}\inst{\ref{aff19},\ref{aff18}}
\and M.~Bolzonella\orcid{0000-0003-3278-4607}\inst{\ref{aff14}}
\and E.~Branchini\orcid{0000-0002-0808-6908}\inst{\ref{aff29},\ref{aff30},\ref{aff25}}
\and M.~Brescia\orcid{0000-0001-9506-5680}\inst{\ref{aff31},\ref{aff16}}
\and S.~Camera\orcid{0000-0003-3399-3574}\inst{\ref{aff32},\ref{aff33},\ref{aff34}}
\and G.~Ca\~nas-Herrera\orcid{0000-0003-2796-2149}\inst{\ref{aff35},\ref{aff36}}
\and V.~Capobianco\orcid{0000-0002-3309-7692}\inst{\ref{aff34}}
\and C.~Carbone\orcid{0000-0003-0125-3563}\inst{\ref{aff37}}
\and J.~Carretero\orcid{0000-0002-3130-0204}\inst{\ref{aff38},\ref{aff39}}
\and S.~Casas\orcid{0000-0002-4751-5138}\inst{\ref{aff40},\ref{aff41}}
\and M.~Castellano\orcid{0000-0001-9875-8263}\inst{\ref{aff42}}
\and G.~Castignani\orcid{0000-0001-6831-0687}\inst{\ref{aff14}}
\and S.~Cavuoti\orcid{0000-0002-3787-4196}\inst{\ref{aff16},\ref{aff43}}
\and K.~C.~Chambers\orcid{0000-0001-6965-7789}\inst{\ref{aff44}}
\and A.~Cimatti\inst{\ref{aff45}}
\and C.~Colodro-Conde\inst{\ref{aff46}}
\and G.~Congedo\orcid{0000-0003-2508-0046}\inst{\ref{aff35}}
\and C.~J.~Conselice\orcid{0000-0003-1949-7638}\inst{\ref{aff47}}
\and L.~Conversi\orcid{0000-0002-6710-8476}\inst{\ref{aff48},\ref{aff24}}
\and Y.~Copin\orcid{0000-0002-5317-7518}\inst{\ref{aff49}}
\and F.~Courbin\orcid{0000-0003-0758-6510}\inst{\ref{aff50},\ref{aff51},\ref{aff52}}
\and H.~M.~Courtois\orcid{0000-0003-0509-1776}\inst{\ref{aff53}}
\and M.~Cropper\orcid{0000-0003-4571-9468}\inst{\ref{aff54}}
\and J.-C.~Cuillandre\orcid{0000-0002-3263-8645}\inst{\ref{aff55}}
\and A.~Da~Silva\orcid{0000-0002-6385-1609}\inst{\ref{aff56},\ref{aff57}}
\and H.~Degaudenzi\orcid{0000-0002-5887-6799}\inst{\ref{aff58}}
\and G.~De~Lucia\orcid{0000-0002-6220-9104}\inst{\ref{aff19}}
\and C.~Dolding\orcid{0009-0003-7199-6108}\inst{\ref{aff54}}
\and H.~Dole\orcid{0000-0002-9767-3839}\inst{\ref{aff23}}
\and F.~Dubath\orcid{0000-0002-6533-2810}\inst{\ref{aff58}}
\and X.~Dupac\inst{\ref{aff24}}
\and S.~Escoffier\orcid{0000-0002-2847-7498}\inst{\ref{aff59}}
\and M.~Farina\orcid{0000-0002-3089-7846}\inst{\ref{aff60}}
\and R.~Farinelli\inst{\ref{aff14}}
\and S.~Ferriol\inst{\ref{aff49}}
\and F.~Finelli\orcid{0000-0002-6694-3269}\inst{\ref{aff14},\ref{aff61}}
\and P.~Fosalba\orcid{0000-0002-1510-5214}\inst{\ref{aff62},\ref{aff12}}
\and S.~Fotopoulou\orcid{0000-0002-9686-254X}\inst{\ref{aff63}}
\and M.~Frailis\orcid{0000-0002-7400-2135}\inst{\ref{aff19}}
\and E.~Franceschi\orcid{0000-0002-0585-6591}\inst{\ref{aff14}}
\and M.~Fumana\orcid{0000-0001-6787-5950}\inst{\ref{aff37}}
\and L.~Gabarra\orcid{0000-0002-8486-8856}\inst{\ref{aff64}}
\and S.~Galeotta\orcid{0000-0002-3748-5115}\inst{\ref{aff19}}
\and K.~George\orcid{0000-0002-1734-8455}\inst{\ref{aff6}}
\and B.~Gillis\orcid{0000-0002-4478-1270}\inst{\ref{aff35}}
\and C.~Giocoli\orcid{0000-0002-9590-7961}\inst{\ref{aff14},\ref{aff15}}
\and P.~G\'omez-Alvarez\orcid{0000-0002-8594-5358}\inst{\ref{aff65},\ref{aff24}}
\and J.~Gracia-Carpio\inst{\ref{aff5}}
\and A.~Grazian\orcid{0000-0002-5688-0663}\inst{\ref{aff66}}
\and F.~Grupp\inst{\ref{aff5},\ref{aff28}}
\and S.~V.~H.~Haugan\orcid{0000-0001-9648-7260}\inst{\ref{aff67}}
\and S.~Hemmati\orcid{0000-0003-2226-5395}\inst{\ref{aff68}}
\and H.~Hoekstra\orcid{0000-0002-0641-3231}\inst{\ref{aff36}}
\and W.~Holmes\inst{\ref{aff69}}
\and F.~Hormuth\inst{\ref{aff70}}
\and A.~Hornstrup\orcid{0000-0002-3363-0936}\inst{\ref{aff71},\ref{aff72}}
\and M.~Huertas-Company\orcid{0000-0002-1416-8483}\inst{\ref{aff46},\ref{aff73},\ref{aff74}}
\and K.~Jahnke\orcid{0000-0003-3804-2137}\inst{\ref{aff75}}
\and M.~Jhabvala\inst{\ref{aff76}}
\and B.~Joachimi\orcid{0000-0001-7494-1303}\inst{\ref{aff77}}
\and S.~Kermiche\orcid{0000-0002-0302-5735}\inst{\ref{aff59}}
\and A.~Kiessling\orcid{0000-0002-2590-1273}\inst{\ref{aff69}}
\and B.~Kubik\orcid{0009-0006-5823-4880}\inst{\ref{aff49}}
\and M.~K\"ummel\orcid{0000-0003-2791-2117}\inst{\ref{aff28}}
\and M.~Kunz\orcid{0000-0002-3052-7394}\inst{\ref{aff78}}
\and H.~Kurki-Suonio\orcid{0000-0002-4618-3063}\inst{\ref{aff79},\ref{aff80}}
\and R.~Laureijs\inst{\ref{aff81}}
\and A.~M.~C.~Le~Brun\orcid{0000-0002-0936-4594}\inst{\ref{aff82}}
\and S.~Ligori\orcid{0000-0003-4172-4606}\inst{\ref{aff34}}
\and P.~B.~Lilje\orcid{0000-0003-4324-7794}\inst{\ref{aff67}}
\and V.~Lindholm\orcid{0000-0003-2317-5471}\inst{\ref{aff79},\ref{aff80}}
\and I.~Lloro\orcid{0000-0001-5966-1434}\inst{\ref{aff83}}
\and M.~Magliocchetti\orcid{0000-0001-9158-4838}\inst{\ref{aff60}}
\and G.~Mainetti\orcid{0000-0003-2384-2377}\inst{\ref{aff84}}
\and D.~Maino\inst{\ref{aff85},\ref{aff37},\ref{aff86}}
\and E.~Maiorano\orcid{0000-0003-2593-4355}\inst{\ref{aff14}}
\and O.~Mansutti\orcid{0000-0001-5758-4658}\inst{\ref{aff19}}
\and S.~Marcin\inst{\ref{aff87}}
\and O.~Marggraf\orcid{0000-0001-7242-3852}\inst{\ref{aff88}}
\and M.~Martinelli\orcid{0000-0002-6943-7732}\inst{\ref{aff42},\ref{aff89}}
\and N.~Martinet\orcid{0000-0003-2786-7790}\inst{\ref{aff90}}
\and F.~Marulli\orcid{0000-0002-8850-0303}\inst{\ref{aff13},\ref{aff14},\ref{aff15}}
\and R.~J.~Massey\orcid{0000-0002-6085-3780}\inst{\ref{aff91}}
\and N.~Mauri\orcid{0000-0001-8196-1548}\inst{\ref{aff45},\ref{aff15}}
\and S.~Maurogordato\inst{\ref{aff92}}
\and E.~Medinaceli\orcid{0000-0002-4040-7783}\inst{\ref{aff14}}
\and S.~Mei\orcid{0000-0002-2849-559X}\inst{\ref{aff93},\ref{aff94}}
\and Y.~Mellier\thanks{Deceased}\inst{\ref{aff7},\ref{aff95}}
\and M.~Meneghetti\orcid{0000-0003-1225-7084}\inst{\ref{aff14},\ref{aff15}}
\and E.~Merlin\orcid{0000-0001-6870-8900}\inst{\ref{aff42}}
\and G.~Meylan\inst{\ref{aff2}}
\and P.~Monaco\orcid{0000-0003-2083-7564}\inst{\ref{aff17},\ref{aff19},\ref{aff20},\ref{aff18}}
\and A.~Mora\orcid{0000-0002-1922-8529}\inst{\ref{aff96}}
\and M.~Moresco\orcid{0000-0002-7616-7136}\inst{\ref{aff13},\ref{aff14}}
\and C.~Moretti\orcid{0000-0003-3314-8936}\inst{\ref{aff19},\ref{aff18},\ref{aff20}}
\and R.~Nakajima\orcid{0009-0009-1213-7040}\inst{\ref{aff88}}
\and C.~Neissner\orcid{0000-0001-8524-4968}\inst{\ref{aff97},\ref{aff39}}
\and S.-M.~Niemi\orcid{0009-0005-0247-0086}\inst{\ref{aff98}}
\and J.~W.~Nightingale\orcid{0000-0002-8987-7401}\inst{\ref{aff99}}
\and C.~Padilla\orcid{0000-0001-7951-0166}\inst{\ref{aff97}}
\and S.~Paltani\orcid{0000-0002-8108-9179}\inst{\ref{aff58}}
\and F.~Pasian\orcid{0000-0002-4869-3227}\inst{\ref{aff19}}
\and K.~Pedersen\inst{\ref{aff100}}
\and W.~J.~Percival\orcid{0000-0002-0644-5727}\inst{\ref{aff3},\ref{aff4},\ref{aff101}}
\and V.~Pettorino\orcid{0000-0002-4203-9320}\inst{\ref{aff98}}
\and A.~Pezzotta\orcid{0000-0003-0726-2268}\inst{\ref{aff25}}
\and S.~Pires\orcid{0000-0002-0249-2104}\inst{\ref{aff55}}
\and G.~Polenta\orcid{0000-0003-4067-9196}\inst{\ref{aff102}}
\and M.~Poncet\inst{\ref{aff103}}
\and L.~A.~Popa\inst{\ref{aff104}}
\and L.~Pozzetti\orcid{0000-0001-7085-0412}\inst{\ref{aff14}}
\and F.~Raison\orcid{0000-0002-7819-6918}\inst{\ref{aff5}}
\and A.~Renzi\orcid{0000-0001-9856-1970}\inst{\ref{aff105},\ref{aff106}}
\and J.~Rhodes\orcid{0000-0002-4485-8549}\inst{\ref{aff69}}
\and G.~Riccio\inst{\ref{aff16}}
\and I.~Risso\orcid{0000-0003-2525-7761}\inst{\ref{aff25},\ref{aff30}}
\and E.~Romelli\orcid{0000-0003-3069-9222}\inst{\ref{aff19}}
\and M.~Roncarelli\orcid{0000-0001-9587-7822}\inst{\ref{aff14}}
\and B.~Rusholme\orcid{0000-0001-7648-4142}\inst{\ref{aff68}}
\and R.~Saglia\orcid{0000-0003-0378-7032}\inst{\ref{aff28},\ref{aff5}}
\and Z.~Sakr\orcid{0000-0002-4823-3757}\inst{\ref{aff10},\ref{aff107},\ref{aff108}}
\and D.~Sapone\orcid{0000-0001-7089-4503}\inst{\ref{aff109}}
\and B.~Sartoris\orcid{0000-0003-1337-5269}\inst{\ref{aff28},\ref{aff19}}
\and M.~Schirmer\orcid{0000-0003-2568-9994}\inst{\ref{aff75}}
\and P.~Schneider\orcid{0000-0001-8561-2679}\inst{\ref{aff88}}
\and T.~Schrabback\orcid{0000-0002-6987-7834}\inst{\ref{aff110}}
\and A.~Secroun\orcid{0000-0003-0505-3710}\inst{\ref{aff59}}
\and E.~Sefusatti\orcid{0000-0003-0473-1567}\inst{\ref{aff19},\ref{aff18},\ref{aff20}}
\and G.~Seidel\orcid{0000-0003-2907-353X}\inst{\ref{aff75}}
\and S.~Serrano\orcid{0000-0002-0211-2861}\inst{\ref{aff62},\ref{aff111},\ref{aff12}}
\and E.~Sihvola\orcid{0000-0003-1804-7715}\inst{\ref{aff112}}
\and P.~Simon\inst{\ref{aff88}}
\and C.~Sirignano\orcid{0000-0002-0995-7146}\inst{\ref{aff105},\ref{aff106}}
\and G.~Sirri\orcid{0000-0003-2626-2853}\inst{\ref{aff15}}
\and J.~Skottfelt\orcid{0000-0003-1310-8283}\inst{\ref{aff113}}
\and L.~Stanco\orcid{0000-0002-9706-5104}\inst{\ref{aff106}}
\and J.~Steinwagner\orcid{0000-0001-7443-1047}\inst{\ref{aff5}}
\and P.~Tallada-Cresp\'{i}\orcid{0000-0002-1336-8328}\inst{\ref{aff38},\ref{aff39}}
\and A.~N.~Taylor\inst{\ref{aff35}}
\and I.~Tereno\orcid{0000-0002-4537-6218}\inst{\ref{aff56},\ref{aff114}}
\and N.~Tessore\orcid{0000-0002-9696-7931}\inst{\ref{aff54}}
\and S.~Toft\orcid{0000-0003-3631-7176}\inst{\ref{aff115},\ref{aff116}}
\and R.~Toledo-Moreo\orcid{0000-0002-2997-4859}\inst{\ref{aff117}}
\and F.~Torradeflot\orcid{0000-0003-1160-1517}\inst{\ref{aff39},\ref{aff38}}
\and I.~Tutusaus\orcid{0000-0002-3199-0399}\inst{\ref{aff12},\ref{aff62},\ref{aff107}}
\and L.~Valenziano\orcid{0000-0002-1170-0104}\inst{\ref{aff14},\ref{aff61}}
\and J.~Valiviita\orcid{0000-0001-6225-3693}\inst{\ref{aff79},\ref{aff80}}
\and T.~Vassallo\orcid{0000-0001-6512-6358}\inst{\ref{aff19}}
\and G.~Verdoes~Kleijn\orcid{0000-0001-5803-2580}\inst{\ref{aff81}}
\and Y.~Wang\orcid{0000-0002-4749-2984}\inst{\ref{aff68}}
\and J.~Weller\orcid{0000-0002-8282-2010}\inst{\ref{aff28},\ref{aff5}}
\and A.~Zacchei\orcid{0000-0003-0396-1192}\inst{\ref{aff19},\ref{aff18}}
\and G.~Zamorani\orcid{0000-0002-2318-301X}\inst{\ref{aff14}}
\and E.~Zucca\orcid{0000-0002-5845-8132}\inst{\ref{aff14}}
\and M.~Ballardini\orcid{0000-0003-4481-3559}\inst{\ref{aff118},\ref{aff119},\ref{aff14}}
\and E.~Bozzo\orcid{0000-0002-8201-1525}\inst{\ref{aff58}}
\and C.~Burigana\orcid{0000-0002-3005-5796}\inst{\ref{aff120},\ref{aff61}}
\and R.~Cabanac\orcid{0000-0001-6679-2600}\inst{\ref{aff107}}
\and M.~Calabrese\orcid{0000-0002-2637-2422}\inst{\ref{aff121},\ref{aff37}}
\and A.~Cappi\inst{\ref{aff92},\ref{aff14}}
\and T.~Castro\orcid{0000-0002-6292-3228}\inst{\ref{aff19},\ref{aff20},\ref{aff18},\ref{aff21}}
\and J.~A.~Escartin~Vigo\inst{\ref{aff5}}
\and J.~Garc\'ia-Bellido\orcid{0000-0002-9370-8360}\inst{\ref{aff122}}
\and V.~Gautard\inst{\ref{aff123}}
\and J.~Macias-Perez\orcid{0000-0002-5385-2763}\inst{\ref{aff124}}
\and R.~Maoli\orcid{0000-0002-6065-3025}\inst{\ref{aff125},\ref{aff42}}
\and J.~Mart\'{i}n-Fleitas\orcid{0000-0002-8594-569X}\inst{\ref{aff126}}
\and R.~B.~Metcalf\orcid{0000-0003-3167-2574}\inst{\ref{aff13},\ref{aff14}}
\and A.~A.~Nucita\inst{\ref{aff127},\ref{aff128},\ref{aff129}}
\and M.~P\"ontinen\orcid{0000-0001-5442-2530}\inst{\ref{aff79}}
\and V.~Scottez\orcid{0009-0008-3864-940X}\inst{\ref{aff7},\ref{aff130}}
\and M.~Sereno\orcid{0000-0003-0302-0325}\inst{\ref{aff14},\ref{aff15}}
\and M.~Tenti\orcid{0000-0002-4254-5901}\inst{\ref{aff15}}
\and M.~Tucci\inst{\ref{aff58}}
\and M.~Viel\orcid{0000-0002-2642-5707}\inst{\ref{aff18},\ref{aff19},\ref{aff26},\ref{aff20},\ref{aff21}}
\and M.~Wiesmann\orcid{0009-0000-8199-5860}\inst{\ref{aff67}}
\and Y.~Akrami\orcid{0000-0002-2407-7956}\inst{\ref{aff122},\ref{aff131}}
\and I.~T.~Andika\orcid{0000-0001-6102-9526}\inst{\ref{aff6},\ref{aff132}}
\and G.~Angora\orcid{0000-0002-0316-6562}\inst{\ref{aff16},\ref{aff118}}
\and S.~Anselmi\orcid{0000-0002-3579-9583}\inst{\ref{aff106},\ref{aff105},\ref{aff133}}
\and M.~Archidiacono\orcid{0000-0003-4952-9012}\inst{\ref{aff85},\ref{aff86}}
\and F.~Atrio-Barandela\orcid{0000-0002-2130-2513}\inst{\ref{aff134}}
\and E.~Aubourg\orcid{0000-0002-5592-023X}\inst{\ref{aff93},\ref{aff135}}
\and L.~Baumont\orcid{0000-0002-1518-0150}\inst{\ref{aff17},\ref{aff19},\ref{aff18}}
\and L.~Bazzanini\orcid{0000-0003-0727-0137}\inst{\ref{aff118},\ref{aff14}}
\and D.~Bertacca\orcid{0000-0002-2490-7139}\inst{\ref{aff105},\ref{aff66},\ref{aff106}}
\and M.~Bethermin\orcid{0000-0002-3915-2015}\inst{\ref{aff22}}
\and F.~Beutler\orcid{0000-0003-0467-5438}\inst{\ref{aff35}}
\and A.~Blanchard\orcid{0000-0001-8555-9003}\inst{\ref{aff107}}
\and L.~Blot\orcid{0000-0002-9622-7167}\inst{\ref{aff136},\ref{aff82}}
\and H.~B\"ohringer\orcid{0000-0001-8241-4204}\inst{\ref{aff5},\ref{aff6},\ref{aff137}}
\and M.~Bonici\orcid{0000-0002-8430-126X}\inst{\ref{aff3},\ref{aff37}}
\and S.~Borgani\orcid{0000-0001-6151-6439}\inst{\ref{aff17},\ref{aff18},\ref{aff19},\ref{aff20},\ref{aff21}}
\and M.~L.~Brown\orcid{0000-0002-0370-8077}\inst{\ref{aff47}}
\and S.~Bruton\orcid{0000-0002-6503-5218}\inst{\ref{aff138}}
\and A.~Calabro\orcid{0000-0003-2536-1614}\inst{\ref{aff42}}
\and B.~Camacho~Quevedo\orcid{0000-0002-8789-4232}\inst{\ref{aff18},\ref{aff26},\ref{aff19}}
\and F.~Caro\inst{\ref{aff42}}
\and C.~S.~Carvalho\inst{\ref{aff114}}
\and Y.~Charles\orcid{0009-0000-3636-834X}\inst{\ref{aff90}}
\and F.~Cogato\orcid{0000-0003-4632-6113}\inst{\ref{aff13},\ref{aff14}}
\and S.~Conseil\orcid{0000-0002-3657-4191}\inst{\ref{aff49}}
\and A.~R.~Cooray\orcid{0000-0002-3892-0190}\inst{\ref{aff139}}
\and M.~Costanzi\orcid{0000-0001-8158-1449}\inst{\ref{aff17},\ref{aff19},\ref{aff18}}
\and O.~Cucciati\orcid{0000-0002-9336-7551}\inst{\ref{aff14}}
\and S.~Davini\orcid{0000-0003-3269-1718}\inst{\ref{aff30}}
\and F.~De~Paolis\orcid{0000-0001-6460-7563}\inst{\ref{aff127},\ref{aff128},\ref{aff129}}
\and G.~Desprez\orcid{0000-0001-8325-1742}\inst{\ref{aff81}}
\and A.~D\'iaz-S\'anchez\orcid{0000-0003-0748-4768}\inst{\ref{aff140}}
\and S.~Di~Domizio\orcid{0000-0003-2863-5895}\inst{\ref{aff29},\ref{aff30}}
\and J.~M.~Diego\orcid{0000-0001-9065-3926}\inst{\ref{aff141}}
\and P.~Dimauro\orcid{0000-0001-7399-2854}\inst{\ref{aff142},\ref{aff42}}
\and P.-A.~Duc\orcid{0000-0003-3343-6284}\inst{\ref{aff22}}
\and V.~Duret\orcid{0009-0009-0383-4960}\inst{\ref{aff59}}
\and M.~Y.~Elkhashab\orcid{0000-0001-9306-2603}\inst{\ref{aff19},\ref{aff20},\ref{aff17},\ref{aff18}}
\and A.~Enia\orcid{0000-0002-0200-2857}\inst{\ref{aff14}}
\and Y.~Fang\orcid{0000-0002-0334-6950}\inst{\ref{aff28}}
\and A.~Finoguenov\orcid{0000-0002-4606-5403}\inst{\ref{aff79}}
\and A.~Franco\orcid{0000-0002-4761-366X}\inst{\ref{aff128},\ref{aff127},\ref{aff129}}
\and K.~Ganga\orcid{0000-0001-8159-8208}\inst{\ref{aff93}}
\and T.~Gasparetto\orcid{0000-0002-7913-4866}\inst{\ref{aff42}}
\and E.~Gaztanaga\orcid{0000-0001-9632-0815}\inst{\ref{aff12},\ref{aff62},\ref{aff143}}
\and F.~Giacomini\orcid{0000-0002-3129-2814}\inst{\ref{aff15}}
\and F.~Gianotti\orcid{0000-0003-4666-119X}\inst{\ref{aff14}}
\and G.~Gozaliasl\orcid{0000-0002-0236-919X}\inst{\ref{aff144},\ref{aff79}}
\and M.~Guidi\orcid{0000-0001-9408-1101}\inst{\ref{aff27},\ref{aff14}}
\and C.~M.~Gutierrez\orcid{0000-0001-7854-783X}\inst{\ref{aff145}}
\and A.~Hall\orcid{0000-0002-3139-8651}\inst{\ref{aff35}}
\and H.~Hildebrandt\orcid{0000-0002-9814-3338}\inst{\ref{aff146}}
\and J.~Hjorth\orcid{0000-0002-4571-2306}\inst{\ref{aff100}}
\and L.~K.~Hunt\orcid{0000-0001-9162-2371}\inst{\ref{aff147}}
\and J.~J.~E.~Kajava\orcid{0000-0002-3010-8333}\inst{\ref{aff148},\ref{aff149},\ref{aff150}}
\and Y.~Kang\orcid{0009-0000-8588-7250}\inst{\ref{aff58}}
\and V.~Kansal\orcid{0000-0002-4008-6078}\inst{\ref{aff151},\ref{aff152}}
\and D.~Karagiannis\orcid{0000-0002-4927-0816}\inst{\ref{aff118},\ref{aff153}}
\and K.~Kiiveri\inst{\ref{aff112}}
\and J.~Kim\orcid{0000-0003-2776-2761}\inst{\ref{aff64}}
\and C.~C.~Kirkpatrick\inst{\ref{aff112}}
\and S.~Kruk\orcid{0000-0001-8010-8879}\inst{\ref{aff24}}
\and M.~Lattanzi\orcid{0000-0003-1059-2532}\inst{\ref{aff119}}
\and L.~Legrand\orcid{0000-0003-0610-5252}\inst{\ref{aff154},\ref{aff155}}
\and F.~Lepori\orcid{0009-0000-5061-7138}\inst{\ref{aff156}}
\and G.~Leroy\orcid{0009-0004-2523-4425}\inst{\ref{aff157},\ref{aff91}}
\and G.~F.~Lesci\orcid{0000-0002-4607-2830}\inst{\ref{aff13},\ref{aff14}}
\and J.~Lesgourgues\orcid{0000-0001-7627-353X}\inst{\ref{aff40}}
\and T.~I.~Liaudat\orcid{0000-0002-9104-314X}\inst{\ref{aff135}}
\and S.~J.~Liu\orcid{0000-0001-7680-2139}\inst{\ref{aff60}}
\and A.~Loureiro\orcid{0000-0002-4371-0876}\inst{\ref{aff158},\ref{aff159}}
\and C.~Mancini\orcid{0000-0002-4297-0561}\inst{\ref{aff37}}
\and F.~Mannucci\orcid{0000-0002-4803-2381}\inst{\ref{aff147}}
\and C.~J.~A.~P.~Martins\orcid{0000-0002-4886-9261}\inst{\ref{aff160},\ref{aff161}}
\and L.~Maurin\orcid{0000-0002-8406-0857}\inst{\ref{aff23}}
\and M.~Miluzio\inst{\ref{aff162},\ref{aff24}}
\and A.~Montoro\orcid{0000-0003-4730-8590}\inst{\ref{aff12},\ref{aff62}}
\and G.~Morgante\inst{\ref{aff14}}
\and S.~Nadathur\orcid{0000-0001-9070-3102}\inst{\ref{aff143}}
\and K.~Naidoo\orcid{0000-0002-9182-1802}\inst{\ref{aff143},\ref{aff75}}
\and P.~Natoli\orcid{0000-0003-0126-9100}\inst{\ref{aff118},\ref{aff119}}
\and A.~Navarro-Alsina\orcid{0000-0002-3173-2592}\inst{\ref{aff88}}
\and S.~Nesseris\orcid{0000-0002-0567-0324}\inst{\ref{aff122}}
\and D.~Paoletti\orcid{0000-0003-4761-6147}\inst{\ref{aff14},\ref{aff61}}
\and F.~Passalacqua\orcid{0000-0002-8606-4093}\inst{\ref{aff105},\ref{aff106}}
\and K.~Paterson\orcid{0000-0001-8340-3486}\inst{\ref{aff75}}
\and L.~Patrizii\inst{\ref{aff15}}
\and A.~Pisani\orcid{0000-0002-6146-4437}\inst{\ref{aff59}}
\and D.~Potter\orcid{0000-0002-0757-5195}\inst{\ref{aff156}}
\and G.~W.~Pratt\inst{\ref{aff55}}
\and S.~Quai\orcid{0000-0002-0449-8163}\inst{\ref{aff13},\ref{aff14}}
\and M.~Radovich\orcid{0000-0002-3585-866X}\inst{\ref{aff66}}
\and G.~Rodighiero\orcid{0000-0002-9415-2296}\inst{\ref{aff105},\ref{aff66}}
\and K.~Rojas\orcid{0000-0003-1391-6854}\inst{\ref{aff87}}
\and W.~Roster\orcid{0000-0002-9149-6528}\inst{\ref{aff5}}
\and S.~Sacquegna\orcid{0000-0002-8433-6630}\inst{\ref{aff163}}
\and M.~Sahl\'en\orcid{0000-0003-0973-4804}\inst{\ref{aff164}}
\and D.~B.~Sanders\orcid{0000-0002-1233-9998}\inst{\ref{aff44}}
\and E.~Sarpa\orcid{0000-0002-1256-655X}\inst{\ref{aff26},\ref{aff21},\ref{aff20}}
\and J.~Schaye\orcid{0000-0002-0668-5560}\inst{\ref{aff36}}
\and A.~Schneider\orcid{0000-0001-7055-8104}\inst{\ref{aff156}}
\and D.~Sciotti\orcid{0009-0008-4519-2620}\inst{\ref{aff42},\ref{aff89}}
\and E.~Sellentin\inst{\ref{aff165},\ref{aff36}}
\and F.~Shankar\orcid{0000-0001-8973-5051}\inst{\ref{aff166}}
\and L.~C.~Smith\orcid{0000-0002-3259-2771}\inst{\ref{aff167}}
\and J.~G.~Sorce\orcid{0000-0002-2307-2432}\inst{\ref{aff168},\ref{aff23}}
\and S.~A.~Stanford\orcid{0000-0003-0122-0841}\inst{\ref{aff169}}
\and K.~Tanidis\orcid{0000-0001-9843-5130}\inst{\ref{aff64}}
\and F.~Tarsitano\orcid{0000-0002-5919-0238}\inst{\ref{aff170},\ref{aff58}}
\and G.~Testera\inst{\ref{aff30}}
\and R.~Teyssier\orcid{0000-0001-7689-0933}\inst{\ref{aff171}}
\and S.~Tosi\orcid{0000-0002-7275-9193}\inst{\ref{aff29},\ref{aff30},\ref{aff25}}
\and A.~Troja\orcid{0000-0003-0239-4595}\inst{\ref{aff105},\ref{aff106}}
\and A.~Venhola\orcid{0000-0001-6071-4564}\inst{\ref{aff172}}
\and D.~Vergani\orcid{0000-0003-0898-2216}\inst{\ref{aff14}}
\and G.~Verza\orcid{0000-0002-1886-8348}$^{\star\star}$\inst{\ref{aff173},\ref{aff174}}
\and P.~Vielzeuf\orcid{0000-0003-2035-9339}\inst{\ref{aff59}}
\and S.~Vinciguerra\orcid{0009-0005-4018-3184}\inst{\ref{aff90}}
\and N.~A.~Walton\orcid{0000-0003-3983-8778}\inst{\ref{aff167}}
\and A.~H.~Wright\orcid{0000-0001-7363-7932}\inst{\ref{aff146}}}
										   
\institute{School of Physics and Astronomy, University of Nottingham, University Park, Nottingham NG7 2RD, UK\label{aff1}
\and
Institute of Physics, Laboratory of Astrophysics, Ecole Polytechnique F\'ed\'erale de Lausanne (EPFL), Observatoire de Sauverny, 1290 Versoix, Switzerland\label{aff2}
\and
Waterloo Centre for Astrophysics, University of Waterloo, Waterloo, Ontario N2L 3G1, Canada\label{aff3}
\and
Department of Physics and Astronomy, University of Waterloo, Waterloo, Ontario N2L 3G1, Canada\label{aff4}
\and
Max Planck Institute for Extraterrestrial Physics, Giessenbachstr. 1, 85748 Garching, Germany\label{aff5}
\and
University Observatory, LMU Faculty of Physics, Scheinerstr.~1, 81679 Munich, Germany\label{aff6}
\and
Institut d'Astrophysique de Paris, 98bis Boulevard Arago, 75014, Paris, France\label{aff7}
\and
OCA, P.H.C Boulevard de l'Observatoire CS 34229, 06304 Nice Cedex 4, France\label{aff8}
\and
Department of Astronomy, University of Florida, Bryant Space Science Center, Gainesville, FL 32611, USA\label{aff9}
\and
Institut f\"ur Theoretische Physik, University of Heidelberg, Philosophenweg 16, 69120 Heidelberg, Germany\label{aff10}
\and
Zentrum f\"ur Astronomie, Universit\"at Heidelberg, Philosophenweg 12, 69120 Heidelberg, Germany\label{aff11}
\and
Institute of Space Sciences (ICE, CSIC), Campus UAB, Carrer de Can Magrans, s/n, 08193 Barcelona, Spain\label{aff12}
\and
Dipartimento di Fisica e Astronomia "Augusto Righi" - Alma Mater Studiorum Universit\`a di Bologna, via Piero Gobetti 93/2, 40129 Bologna, Italy\label{aff13}
\and
INAF-Osservatorio di Astrofisica e Scienza dello Spazio di Bologna, Via Piero Gobetti 93/3, 40129 Bologna, Italy\label{aff14}
\and
INFN-Sezione di Bologna, Viale Berti Pichat 6/2, 40127 Bologna, Italy\label{aff15}
\and
INAF-Osservatorio Astronomico di Capodimonte, Via Moiariello 16, 80131 Napoli, Italy\label{aff16}
\and
Dipartimento di Fisica - Sezione di Astronomia, Universit\`a di Trieste, Via Tiepolo 11, 34131 Trieste, Italy\label{aff17}
\and
IFPU, Institute for Fundamental Physics of the Universe, via Beirut 2, 34151 Trieste, Italy\label{aff18}
\and
INAF-Osservatorio Astronomico di Trieste, Via G. B. Tiepolo 11, 34143 Trieste, Italy\label{aff19}
\and
INFN, Sezione di Trieste, Via Valerio 2, 34127 Trieste TS, Italy\label{aff20}
\and
ICSC - Centro Nazionale di Ricerca in High Performance Computing, Big Data e Quantum Computing, Via Magnanelli 2, Bologna, Italy\label{aff21}
\and
Universit\'e de Strasbourg, CNRS, Observatoire astronomique de Strasbourg, UMR 7550, 67000 Strasbourg, France\label{aff22}
\and
Universit\'e Paris-Saclay, CNRS, Institut d'astrophysique spatiale, 91405, Orsay, France\label{aff23}
\and
ESAC/ESA, Camino Bajo del Castillo, s/n., Urb. Villafranca del Castillo, 28692 Villanueva de la Ca\~nada, Madrid, Spain\label{aff24}
\and
INAF-Osservatorio Astronomico di Brera, Via Brera 28, 20122 Milano, Italy\label{aff25}
\and
SISSA, International School for Advanced Studies, Via Bonomea 265, 34136 Trieste TS, Italy\label{aff26}
\and
Dipartimento di Fisica e Astronomia, Universit\`a di Bologna, Via Gobetti 93/2, 40129 Bologna, Italy\label{aff27}
\and
Universit\"ats-Sternwarte M\"unchen, Fakult\"at f\"ur Physik, Ludwig-Maximilians-Universit\"at M\"unchen, Scheinerstr.~1, 81679 M\"unchen, Germany\label{aff28}
\and
Dipartimento di Fisica, Universit\`a di Genova, Via Dodecaneso 33, 16146, Genova, Italy\label{aff29}
\and
INFN-Sezione di Genova, Via Dodecaneso 33, 16146, Genova, Italy\label{aff30}
\and
Department of Physics "E. Pancini", University Federico II, Via Cinthia 6, 80126, Napoli, Italy\label{aff31}
\and
Dipartimento di Fisica, Universit\`a degli Studi di Torino, Via P. Giuria 1, 10125 Torino, Italy\label{aff32}
\and
INFN-Sezione di Torino, Via P. Giuria 1, 10125 Torino, Italy\label{aff33}
\and
INAF-Osservatorio Astrofisico di Torino, Via Osservatorio 20, 10025 Pino Torinese (TO), Italy\label{aff34}
\and
Institute for Astronomy, University of Edinburgh, Royal Observatory, Blackford Hill, Edinburgh EH9 3HJ, UK\label{aff35}
\and
Leiden Observatory, Leiden University, Einsteinweg 55, 2333 CC Leiden, The Netherlands\label{aff36}
\and
INAF-IASF Milano, Via Alfonso Corti 12, 20133 Milano, Italy\label{aff37}
\and
Centro de Investigaciones Energ\'eticas, Medioambientales y Tecnol\'ogicas (CIEMAT), Avenida Complutense 40, 28040 Madrid, Spain\label{aff38}
\and
Port d'Informaci\'{o} Cient\'{i}fica, Campus UAB, C. Albareda s/n, 08193 Bellaterra (Barcelona), Spain\label{aff39}
\and
Institute for Theoretical Particle Physics and Cosmology (TTK), RWTH Aachen University, 52056 Aachen, Germany\label{aff40}
\and
Deutsches Zentrum f\"ur Luft- und Raumfahrt e. V. (DLR), Linder H\"ohe, 51147 K\"oln, Germany\label{aff41}
\and
INAF-Osservatorio Astronomico di Roma, Via Frascati 33, 00078 Monteporzio Catone, Italy\label{aff42}
\and
INFN -- Sezione di Napoli, Via Cinthia 6, 80126, Napoli, Italy\label{aff43}
\and
Institute for Astronomy, University of Hawaii, 2680 Woodlawn Drive, Honolulu, HI 96822, USA\label{aff44}
\and
Dipartimento di Fisica e Astronomia "Augusto Righi" - Alma Mater Studiorum Universit\`a di Bologna, Viale Berti Pichat 6/2, 40127 Bologna, Italy\label{aff45}
\and
Instituto de Astrof\'{\i}sica de Canarias, E-38205 La Laguna, Tenerife, Spain\label{aff46}
\and
Jodrell Bank Centre for Astrophysics, Department of Physics and Astronomy, University of Manchester, Oxford Road, Manchester M13 9PL, UK\label{aff47}
\and
European Space Agency/ESRIN, Largo Galileo Galilei 1, 00044 Frascati, Roma, Italy\label{aff48}
\and
Universit\'e Claude Bernard Lyon 1, CNRS/IN2P3, IP2I Lyon, UMR 5822, Villeurbanne, F-69100, France\label{aff49}
\and
Institut de Ci\`{e}ncies del Cosmos (ICCUB), Universitat de Barcelona (IEEC-UB), Mart\'{i} i Franqu\`{e}s 1, 08028 Barcelona, Spain\label{aff50}
\and
Instituci\'o Catalana de Recerca i Estudis Avan\c{c}ats (ICREA), Passeig de Llu\'{\i}s Companys 23, 08010 Barcelona, Spain\label{aff51}
\and
Institut de Ciencies de l'Espai (IEEC-CSIC), Campus UAB, Carrer de Can Magrans, s/n Cerdanyola del Vall\'es, 08193 Barcelona, Spain\label{aff52}
\and
UCB Lyon 1, CNRS/IN2P3, IUF, IP2I Lyon, 4 rue Enrico Fermi, 69622 Villeurbanne, France\label{aff53}
\and
Mullard Space Science Laboratory, University College London, Holmbury St Mary, Dorking, Surrey RH5 6NT, UK\label{aff54}
\and
Universit\'e Paris-Saclay, Universit\'e Paris Cit\'e, CEA, CNRS, AIM, 91191, Gif-sur-Yvette, France\label{aff55}
\and
Departamento de F\'isica, Faculdade de Ci\^encias, Universidade de Lisboa, Edif\'icio C8, Campo Grande, PT1749-016 Lisboa, Portugal\label{aff56}
\and
Instituto de Astrof\'isica e Ci\^encias do Espa\c{c}o, Faculdade de Ci\^encias, Universidade de Lisboa, Campo Grande, 1749-016 Lisboa, Portugal\label{aff57}
\and
Department of Astronomy, University of Geneva, ch. d'Ecogia 16, 1290 Versoix, Switzerland\label{aff58}
\and
Aix-Marseille Universit\'e, CNRS/IN2P3, CPPM, Marseille, France\label{aff59}
\and
INAF-Istituto di Astrofisica e Planetologia Spaziali, via del Fosso del Cavaliere, 100, 00100 Roma, Italy\label{aff60}
\and
INFN-Bologna, Via Irnerio 46, 40126 Bologna, Italy\label{aff61}
\and
Institut d'Estudis Espacials de Catalunya (IEEC),  Edifici RDIT, Campus UPC, 08860 Castelldefels, Barcelona, Spain\label{aff62}
\and
School of Physics, HH Wills Physics Laboratory, University of Bristol, Tyndall Avenue, Bristol, BS8 1TL, UK\label{aff63}
\and
Department of Physics, University of Oxford, Keble Road, Oxford OX1 3RH, UK\label{aff64}
\and
FRACTAL S.L.N.E., calle Tulip\'an 2, Portal 13 1A, 28231, Las Rozas de Madrid, Spain\label{aff65}
\and
INAF-Osservatorio Astronomico di Padova, Via dell'Osservatorio 5, 35122 Padova, Italy\label{aff66}
\and
Institute of Theoretical Astrophysics, University of Oslo, P.O. Box 1029 Blindern, 0315 Oslo, Norway\label{aff67}
\and
Caltech/IPAC, 1200 E. California Blvd., Pasadena, CA 91125, USA\label{aff68}
\and
Jet Propulsion Laboratory, California Institute of Technology, 4800 Oak Grove Drive, Pasadena, CA, 91109, USA\label{aff69}
\and
Felix Hormuth Engineering, Goethestr. 17, 69181 Leimen, Germany\label{aff70}
\and
Technical University of Denmark, Elektrovej 327, 2800 Kgs. Lyngby, Denmark\label{aff71}
\and
Cosmic Dawn Center (DAWN), Denmark\label{aff72}
\and
Universit\'e PSL, Observatoire de Paris, Sorbonne Universit\'e, CNRS, LERMA, 75014, Paris, France\label{aff73}
\and
Universit\'e Paris-Cit\'e, 5 Rue Thomas Mann, 75013, Paris, France\label{aff74}
\and
Max-Planck-Institut f\"ur Astronomie, K\"onigstuhl 17, 69117 Heidelberg, Germany\label{aff75}
\and
NASA Goddard Space Flight Center, Greenbelt, MD 20771, USA\label{aff76}
\and
Department of Physics and Astronomy, University College London, Gower Street, London WC1E 6BT, UK\label{aff77}
\and
Universit\'e de Gen\`eve, D\'epartement de Physique Th\'eorique and Centre for Astroparticle Physics, 24 quai Ernest-Ansermet, CH-1211 Gen\`eve 4, Switzerland\label{aff78}
\and
Department of Physics, P.O. Box 64, University of Helsinki, 00014 Helsinki, Finland\label{aff79}
\and
Helsinki Institute of Physics, Gustaf H{\"a}llstr{\"o}min katu 2, University of Helsinki, 00014 Helsinki, Finland\label{aff80}
\and
Kapteyn Astronomical Institute, University of Groningen, PO Box 800, 9700 AV Groningen, The Netherlands\label{aff81}
\and
Laboratoire d'etude de l'Univers et des phenomenes eXtremes, Observatoire de Paris, Universit\'e PSL, Sorbonne Universit\'e, CNRS, 92190 Meudon, France\label{aff82}
\and
SKAO, Jodrell Bank, Lower Withington, Macclesfield SK11 9FT, UK\label{aff83}
\and
Centre de Calcul de l'IN2P3/CNRS, 21 avenue Pierre de Coubertin 69627 Villeurbanne Cedex, France\label{aff84}
\and
Dipartimento di Fisica "Aldo Pontremoli", Universit\`a degli Studi di Milano, Via Celoria 16, 20133 Milano, Italy\label{aff85}
\and
INFN-Sezione di Milano, Via Celoria 16, 20133 Milano, Italy\label{aff86}
\and
University of Applied Sciences and Arts of Northwestern Switzerland, School of Computer Science, 5210 Windisch, Switzerland\label{aff87}
\and
Universit\"at Bonn, Argelander-Institut f\"ur Astronomie, Auf dem H\"ugel 71, 53121 Bonn, Germany\label{aff88}
\and
INFN-Sezione di Roma, Piazzale Aldo Moro, 2 - c/o Dipartimento di Fisica, Edificio G. Marconi, 00185 Roma, Italy\label{aff89}
\and
Aix-Marseille Universit\'e, CNRS, CNES, LAM, Marseille, France\label{aff90}
\and
Department of Physics, Institute for Computational Cosmology, Durham University, South Road, Durham, DH1 3LE, UK\label{aff91}
\and
Universit\'e C\^{o}te d'Azur, Observatoire de la C\^{o}te d'Azur, CNRS, Laboratoire Lagrange, Bd de l'Observatoire, CS 34229, 06304 Nice cedex 4, France\label{aff92}
\and
Universit\'e Paris Cit\'e, CNRS, Astroparticule et Cosmologie, 75013 Paris, France\label{aff93}
\and
CNRS-UCB International Research Laboratory, Centre Pierre Bin\'etruy, IRL2007, CPB-IN2P3, Berkeley, USA\label{aff94}
\and
Institut d'Astrophysique de Paris, UMR 7095, CNRS, and Sorbonne Universit\'e, 98 bis boulevard Arago, 75014 Paris, France\label{aff95}
\and
Telespazio UK S.L. for European Space Agency (ESA), Camino bajo del Castillo, s/n, Urbanizacion Villafranca del Castillo, Villanueva de la Ca\~nada, 28692 Madrid, Spain\label{aff96}
\and
Institut de F\'{i}sica d'Altes Energies (IFAE), The Barcelona Institute of Science and Technology, Campus UAB, 08193 Bellaterra (Barcelona), Spain\label{aff97}
\and
European Space Agency/ESTEC, Keplerlaan 1, 2201 AZ Noordwijk, The Netherlands\label{aff98}
\and
School of Mathematics, Statistics and Physics, Newcastle University, Herschel Building, Newcastle-upon-Tyne, NE1 7RU, UK\label{aff99}
\and
DARK, Niels Bohr Institute, University of Copenhagen, Jagtvej 155, 2200 Copenhagen, Denmark\label{aff100}
\and
Perimeter Institute for Theoretical Physics, Waterloo, Ontario N2L 2Y5, Canada\label{aff101}
\and
Space Science Data Center, Italian Space Agency, via del Politecnico snc, 00133 Roma, Italy\label{aff102}
\and
Centre National d'Etudes Spatiales -- Centre spatial de Toulouse, 18 avenue Edouard Belin, 31401 Toulouse Cedex 9, France\label{aff103}
\and
Institute of Space Science, Str. Atomistilor, nr. 409 M\u{a}gurele, Ilfov, 077125, Romania\label{aff104}
\and
Dipartimento di Fisica e Astronomia "G. Galilei", Universit\`a di Padova, Via Marzolo 8, 35131 Padova, Italy\label{aff105}
\and
INFN-Padova, Via Marzolo 8, 35131 Padova, Italy\label{aff106}
\and
Institut de Recherche en Astrophysique et Plan\'etologie (IRAP), Universit\'e de Toulouse, CNRS, UPS, CNES, 14 Av. Edouard Belin, 31400 Toulouse, France\label{aff107}
\and
Universit\'e St Joseph; Faculty of Sciences, Beirut, Lebanon\label{aff108}
\and
Departamento de F\'isica, FCFM, Universidad de Chile, Blanco Encalada 2008, Santiago, Chile\label{aff109}
\and
Universit\"at Innsbruck, Institut f\"ur Astro- und Teilchenphysik, Technikerstr. 25/8, 6020 Innsbruck, Austria\label{aff110}
\and
Satlantis, University Science Park, Sede Bld 48940, Leioa-Bilbao, Spain\label{aff111}
\and
Department of Physics and Helsinki Institute of Physics, Gustaf H\"allstr\"omin katu 2, University of Helsinki, 00014 Helsinki, Finland\label{aff112}
\and
Centre for Electronic Imaging, Open University, Walton Hall, Milton Keynes, MK7~6AA, UK\label{aff113}
\and
Instituto de Astrof\'isica e Ci\^encias do Espa\c{c}o, Faculdade de Ci\^encias, Universidade de Lisboa, Tapada da Ajuda, 1349-018 Lisboa, Portugal\label{aff114}
\and
Cosmic Dawn Center (DAWN)\label{aff115}
\and
Niels Bohr Institute, University of Copenhagen, Jagtvej 128, 2200 Copenhagen, Denmark\label{aff116}
\and
Universidad Polit\'ecnica de Cartagena, Departamento de Electr\'onica y Tecnolog\'ia de Computadoras,  Plaza del Hospital 1, 30202 Cartagena, Spain\label{aff117}
\and
Dipartimento di Fisica e Scienze della Terra, Universit\`a degli Studi di Ferrara, Via Giuseppe Saragat 1, 44122 Ferrara, Italy\label{aff118}
\and
Istituto Nazionale di Fisica Nucleare, Sezione di Ferrara, Via Giuseppe Saragat 1, 44122 Ferrara, Italy\label{aff119}
\and
INAF, Istituto di Radioastronomia, Via Piero Gobetti 101, 40129 Bologna, Italy\label{aff120}
\and
Astronomical Observatory of the Autonomous Region of the Aosta Valley (OAVdA), Loc. Lignan 39, I-11020, Nus (Aosta Valley), Italy\label{aff121}
\and
Instituto de F\'isica Te\'orica UAM-CSIC, Campus de Cantoblanco, 28049 Madrid, Spain\label{aff122}
\and
CEA Saclay, DFR/IRFU, Service d'Astrophysique, Bat. 709, 91191 Gif-sur-Yvette, France\label{aff123}
\and
Univ. Grenoble Alpes, CNRS, Grenoble INP, LPSC-IN2P3, 53, Avenue des Martyrs, 38000, Grenoble, France\label{aff124}
\and
Dipartimento di Fisica, Sapienza Universit\`a di Roma, Piazzale Aldo Moro 2, 00185 Roma, Italy\label{aff125}
\and
Aurora Technology for European Space Agency (ESA), Camino bajo del Castillo, s/n, Urbanizacion Villafranca del Castillo, Villanueva de la Ca\~nada, 28692 Madrid, Spain\label{aff126}
\and
Department of Mathematics and Physics E. De Giorgi, University of Salento, Via per Arnesano, CP-I93, 73100, Lecce, Italy\label{aff127}
\and
INFN, Sezione di Lecce, Via per Arnesano, CP-193, 73100, Lecce, Italy\label{aff128}
\and
INAF-Sezione di Lecce, c/o Dipartimento Matematica e Fisica, Via per Arnesano, 73100, Lecce, Italy\label{aff129}
\and
ICL, Junia, Universit\'e Catholique de Lille, LITL, 59000 Lille, France\label{aff130}
\and
CERCA/ISO, Department of Physics, Case Western Reserve University, 10900 Euclid Avenue, Cleveland, OH 44106, USA\label{aff131}
\and
Technical University of Munich, TUM School of Natural Sciences, Physics Department, James-Franck-Str.~1, 85748 Garching, Germany\label{aff132}
\and
Laboratoire Univers et Th\'eorie, Observatoire de Paris, Universit\'e PSL, Universit\'e Paris Cit\'e, CNRS, 92190 Meudon, France\label{aff133}
\and
Departamento de F{\'\i}sica Fundamental. Universidad de Salamanca. Plaza de la Merced s/n. 37008 Salamanca, Spain\label{aff134}
\and
IRFU, CEA, Universit\'e Paris-Saclay 91191 Gif-sur-Yvette Cedex, France\label{aff135}
\and
Center for Data-Driven Discovery, Kavli IPMU (WPI), UTIAS, The University of Tokyo, Kashiwa, Chiba 277-8583, Japan\label{aff136}
\and
Max-Planck-Institut f\"ur Physik, Boltzmannstr. 8, 85748 Garching, Germany\label{aff137}
\and
California Institute of Technology, 1200 E California Blvd, Pasadena, CA 91125, USA\label{aff138}
\and
Department of Physics \& Astronomy, University of California Irvine, Irvine CA 92697, USA\label{aff139}
\and
Departamento F\'isica Aplicada, Universidad Polit\'ecnica de Cartagena, Campus Muralla del Mar, 30202 Cartagena, Murcia, Spain\label{aff140}
\and
Instituto de F\'isica de Cantabria, Edificio Juan Jord\'a, Avenida de los Castros, 39005 Santander, Spain\label{aff141}
\and
Observatorio Nacional, Rua General Jose Cristino, 77-Bairro Imperial de Sao Cristovao, Rio de Janeiro, 20921-400, Brazil\label{aff142}
\and
Institute of Cosmology and Gravitation, University of Portsmouth, Portsmouth PO1 3FX, UK\label{aff143}
\and
Department of Computer Science, Aalto University, PO Box 15400, Espoo, FI-00 076, Finland\label{aff144}
\and
 Instituto de Astrof\'{\i}sica de Canarias, E-38205 La Laguna; Universidad de La Laguna, Dpto. Astrof\'\i sica, E-38206 La Laguna, Tenerife, Spain\label{aff145}
\and
Ruhr University Bochum, Faculty of Physics and Astronomy, Astronomical Institute (AIRUB), German Centre for Cosmological Lensing (GCCL), 44780 Bochum, Germany\label{aff146}
\and
INAF-Osservatorio Astrofisico di Arcetri, Largo E. Fermi 5, 50125, Firenze, Italy\label{aff147}
\and
Department of Physics and Astronomy, Vesilinnantie 5, University of Turku, 20014 Turku, Finland\label{aff148}
\and
Finnish Centre for Astronomy with ESO (FINCA), Quantum, Vesilinnantie 5, University of Turku, 20014 Turku, Finland\label{aff149}
\and
Serco for European Space Agency (ESA), Camino bajo del Castillo, s/n, Urbanizacion Villafranca del Castillo, Villanueva de la Ca\~nada, 28692 Madrid, Spain\label{aff150}
\and
ARC Centre of Excellence for Dark Matter Particle Physics, Melbourne, Australia\label{aff151}
\and
Centre for Astrophysics \& Supercomputing, Swinburne University of Technology,  Hawthorn, Victoria 3122, Australia\label{aff152}
\and
Department of Physics and Astronomy, University of the Western Cape, Bellville, Cape Town, 7535, South Africa\label{aff153}
\and
DAMTP, Centre for Mathematical Sciences, Wilberforce Road, Cambridge CB3 0WA, UK\label{aff154}
\and
Kavli Institute for Cosmology Cambridge, Madingley Road, Cambridge, CB3 0HA, UK\label{aff155}
\and
Department of Astrophysics, University of Zurich, Winterthurerstrasse 190, 8057 Zurich, Switzerland\label{aff156}
\and
Department of Physics, Centre for Extragalactic Astronomy, Durham University, South Road, Durham, DH1 3LE, UK\label{aff157}
\and
Oskar Klein Centre for Cosmoparticle Physics, Department of Physics, Stockholm University, Stockholm, SE-106 91, Sweden\label{aff158}
\and
Astrophysics Group, Blackett Laboratory, Imperial College London, London SW7 2AZ, UK\label{aff159}
\and
Centro de Astrof\'{\i}sica da Universidade do Porto, Rua das Estrelas, 4150-762 Porto, Portugal\label{aff160}
\and
Instituto de Astrof\'isica e Ci\^encias do Espa\c{c}o, Universidade do Porto, CAUP, Rua das Estrelas, PT4150-762 Porto, Portugal\label{aff161}
\and
HE Space for European Space Agency (ESA), Camino bajo del Castillo, s/n, Urbanizacion Villafranca del Castillo, Villanueva de la Ca\~nada, 28692 Madrid, Spain\label{aff162}
\and
INAF - Osservatorio Astronomico d'Abruzzo, Via Maggini, 64100, Teramo, Italy\label{aff163}
\and
Theoretical astrophysics, Department of Physics and Astronomy, Uppsala University, Box 516, 751 37 Uppsala, Sweden\label{aff164}
\and
Mathematical Institute, University of Leiden, Einsteinweg 55, 2333 CA Leiden, The Netherlands\label{aff165}
\and
School of Physics \& Astronomy, University of Southampton, Highfield Campus, Southampton SO17 1BJ, UK\label{aff166}
\and
Institute of Astronomy, University of Cambridge, Madingley Road, Cambridge CB3 0HA, UK\label{aff167}
\and
Univ. Lille, CNRS, Centrale Lille, UMR 9189 CRIStAL, 59000 Lille, France\label{aff168}
\and
Department of Physics and Astronomy, University of California, Davis, CA 95616, USA\label{aff169}
\and
Institute for Particle Physics and Astrophysics, Dept. of Physics, ETH Zurich, Wolfgang-Pauli-Strasse 27, 8093 Zurich, Switzerland\label{aff170}
\and
Department of Astrophysical Sciences, Peyton Hall, Princeton University, Princeton, NJ 08544, USA\label{aff171}
\and
Space physics and astronomy research unit, University of Oulu, Pentti Kaiteran katu 1, FI-90014 Oulu, Finland\label{aff172}
\and
International Centre for Theoretical Physics (ICTP), Strada Costiera 11, 34151 Trieste, Italy\label{aff173}
\and
Center for Computational Astrophysics, Flatiron Institute, 162 5th Avenue, 10010, New York, NY, USA\label{aff174}}    

\date{Received 30 March 2026 / Accepted 21 July 2026}

\abstract{Intracluster light (ICL) offers a luminous route for measuring the shapes and orientations of cluster haloes. We used Euclid Quick Data Release (Q1) images to quantify ICL shapes for 205 clusters at $0.1 \leq z \leq 0.8$ selected from the eROSITA All-Sky Survey and the Dark Energy Survey. We measured the ICL properties across five photometric bands (\IE, \YE, \JE, \HE and the coadded $YJ\HE$), and we find that the shape parameters are consistent across the filters, with the \HE band probing the profiles to the largest cluster-centric distances ($0.1\,R_{200\mathrm{c}}$ $\sim$\,100\,kpc). The ICL ellipticity ($e=1-b/a$) distribution peaks at 0.5, in close agreement with halo ellipticities inferred from strong and weak lensing studies, suggesting that the distribution of ICL traces the large-scale structure of the clusters. Cluster cores, which are dominated by brightest cluster galaxy (BCG) light, are round ($e\sim$\,0.2), while the ICL becomes progressively more elongated at large radii, reaching an ellipticity of $\sim$\,0.5 by $0.1\,R_{200\mathrm{c}}$. 
We find that the ICL is closely aligned with the BCG (87\% lie within 30$^\circ$), but it exhibits a weaker alignment with the satellite galaxy distribution (61\% within 30$^\circ$). Furthermore, robust position angles could be measured using the ICL for 63\% of the sample, whereas position angles could be measured for only 30\% of the sample using the satellite galaxies. Stacking clusters in redshift bins allows for measurements of the ICL surface brightness to $0.6$--$0.8\,R_{200\mathrm{c}}$ and ellipticity beyond $0.3$--$0.4\,R_{200\mathrm{c}}$ (300--400\,kpc), but we observed a plateau in the ICL ellipticity beyond $0.1\,R_{200\mathrm{c}}$. These results demonstrate the capability of \Euclid to characterise the ICL morphology with a high statistical precision, which will allow us to place new constraints on cluster halo structure and its evolution with redshift.
}

   \titlerunning{Shape of cluster haloes revealed by ICL}
   \authorrunning{Euclid Collaboration: T. Kolcu et al.}
   
   \maketitle 

\section{Introduction}
In the current cosmological paradigm, the majority of matter in the Universe exists in the form of cold dark matter (CDM) -- a non-relativistic, collisionless component that interacts primarily through gravity. Simulations predict that in the standard CDM model, the collisionless and anisotropic nature of dark matter collapse leads to the formation of triaxial, typically prolate, haloes \citep{Dubinski_1991, Jing_2002, Oguri_2005,Allgood_2006,Munoz_Cuartas_2011,Vega-Ferrero_2017,Cataldi_2023}. Measurements of dark matter halo shapes can provide information about how they assemble and their relation to the ambient tidal field. Shapes are also sensitive to dark matter microphysics. Alternative dark matter models, such as warm dark matter or self-interacting dark matter, result in rounder and more spherical halo shapes \citep{Miralda_Escude_2002, Brinckmann_2018, Despali_2022, Gonzalez_2024}, so robust constraints on halo ellipticity translate into bounds on deviations from collisionless CDM. 

At the cluster scale, a triaxial structure inflates projected concentrations of galaxies and lensing cross-sections, thereby biasing weak lensing mass inferences and the effective selection function of the clusters. This means that halo triaxiality is a major source of intrinsic scatter in weak gravitational lensing mass measurements of clusters 
\citep[e.g.][]{Meneghetti_2010,Becker_Kravtsov_2011,Giocoli_2014,Meneghetti_2014}. These systematic errors propagate through mass-scaling relation calibration and cluster-abundance analyses to shift constraints on the matter density parameter, $\Omega_{\rm m}$, and the amplitude of matter fluctuations on $8\,h^{-1}\,\mathrm{Mpc}$ scales, $\sigma_{8}$. \citep[e.g.][]{Oguri_2005,Corless_King_2007,Meneghetti_2010,Umetsu_2022}. Finally, mapping how the halo shape changes with radius can constrain how baryons reshape the potential (via adiabatic contraction or feedback-driven heating; \citealt{Gnedin_2004}), offering an independent test of galaxy-formation physics \citep{Kazantzidis_2004,Abadi_2010,Bryan_2013,Velliscig_2015,Chua_2019}. This motivates precise radius-dependent measurements of cluster halo shapes that can serve as a test of the $\Lambda$CDM, enable the calibration of mass-observable relations, and provide information on galaxy formation physics.

Since dark matter is non-luminous, its halo morphology must be inferred indirectly using luminous tracers. One widely used method is gravitational lensing of background galaxies \citep{Grossman_1989,Bartelmann_2010}, which enables reconstruction of the mass distribution in galaxy clusters \citep[e.g.][]{Kneib_2011, Knebe_2013} and can be used to estimate halo ellipticities \citep[e.g.][]{Evans_2009, Oguri_2010, vanUitert_2017, Shin_2018}. Another approach involves using satellite galaxies \citep{Herle_2025} or the intracluster medium (ICM) of galaxy clusters to trace the underlying gravitational potential and halo shape \citep{Eyles_1991,Clowe_2004,Montes_2019}.

While the methods discussed above provide complementary measurements of the shape of dark matter haloes, each is subject to specific limitations. Gravitational lensing measurements can be affected by systematic observational uncertainties, such as contamination from foreground or background sources, misidentification of multiply lensed images, and the necessity of stacking to enhance intrinsically weak signals (from weak gravitational lensing). Satellite galaxy positions are often limited by the small number of member galaxies in individual clusters \citep{Herle_2025}, resulting in noisy data that again require stacking across multiple systems. On the other hand, the ICM is collisional and may not trace the collisionless dark matter distribution well, particularly in dynamically disturbed or merging clusters (e.g. \citealt{Clowe_2004}; \citealt{Markevitch_2004}; \citealt{Montes_2019}).

Over the past decade, both observations and simulations of intracluster light (ICL) have shown that it acts as a reliable tracer of dark matter haloes in galaxy clusters \citep{Montes_2019,AlonsoAsensio_2020,Contreras-Santos_24,Yoo_2024,Butler_2025,Ellien25}. The ICL consists of diffuse light from stars that are gravitationally bound to the cluster potential rather than to any individual galaxy, and it is therefore expected to trace the gravitational potential of the underlying dark matter halo (see the review by \citealt{Montes_2022}). These intracluster stars arise primarily from the tidal stripping of galaxies \citep{Gregg_West_1998,Purcell_2007,Martin_2024, Spavone_2024} as well as major mergers, often with the brightest cluster galaxy (BCG; \citealt{Contini_2014, Chun_2024}) and preprocessing in infalling groups \citep{Rudick_2006,Murante_2007,Contini_2014, Spavone_2018}. Because the intracluster stars are bound to the underlying dark matter distribution, measuring the spatial and morphological properties of the ICL can offer an indirect means of probing the shape of dark matter haloes on an individual cluster basis.

From an observational standpoint, characterising ICL has historically been challenging due to its faint surface brightness, for example $\mu_r(3\sigma,10^{\arcsec}\times10^{\arcsec})>26\,\mathrm{mag\,arcsec^{-2}}$ \citep{Gonzalez_2005,Mihos_2005,Rudick_2006,Montes_2021,Brough_2024}, and extended (hundreds of kiloparsecs) nature, which requires deep and wide-field observations. In recent years, with deep optical imaging and careful background subtraction techniques developed, the ICL has been robustly characterised in significant numbers of clusters. Large surveys -- for example, the Dark Energy Survey (\citealt{DES_2005}; e.g. \citealt{Zhang_2019,Golden-Marx_2023,Golden-Marx_2025}) and the Hyper Suprime-Cam Subaru Strategic Programme \citep{Aihara_2018,Aihara_2022} -- have enabled studies of ICL in hundreds of clusters. These observations have revealed clear connections between the ICL and cluster properties. For instance, ICL (plus BCG) luminosity correlates with cluster mass, size, and velocity dispersion, indicating that more massive clusters tend to exhibit more extended diffuse light \citep{Kluge_2021,Ragusa_2021,Golden-Marx_2023,Contini_2024_vegas,Golden-Marx_2025}. 

The \Euclid space mission \citep{Laureijs11, EuclidSkyOverview} is designed to map approximately one third of the sky (${\sim} 14\,000$ square degrees) with exceptional sensitivity and image quality using four broad photometric filters. Due to its faint detection limits, wide field of view, and tight control of scattered light, \Euclid is well suited for studying the low surface brightness Universe \citep{Scaramella-EP1, Borlaff-EP16}, particularly ICL \citep{EP-Bellhouse}. This capability has already been demonstrated through Euclid Early Release Observations, which reveal the ICL and other faint diffuse structures with unprecedented clarity \citep{EROData, EROPerseusICL, Saifollahi25, Urbano24}. In this study, we use data from the Euclid Quick Data Release \citep[Q1, ][]{Q1cite} to characterise the halo shapes of the ICL in 205 galaxy clusters. Taking advantage of the multi-band imaging of \Euclid and deep observational capabilities, we aim to overcome challenges in detecting and measuring the morphology of the ICL in clusters. The methodology developed here will be extended to the complete \Euclid survey in future data releases, allowing a statistically robust analysis of ICL shape and its dependence on halo mass and redshift across thousands of systems.

The structure of this paper is as follows. In Sect.~\ref{sec:observations_sample} we describe the Euclid Q1 observations and introduce the cluster sample used in this work. We also describe our methodology for extracting the radial surface brightness and ICL halo shape profiles. In Sect.~\ref{sec:Euclid_filter_sensitivity_to_ICL} we compare the sensitivity of the different photometric filters that are used to trace the ICL signal and its shape. Section~\ref{sec:BCG_ICL_ellipticity_distribution} presents the ellipticity distribution of the ICL alongside comparisons with ellipticities inferred from lensing studies and Hydrangea hydrodynamical simulations. In Sect.~\ref{sec:average_shape_profiles} we discuss the variation in ICL halo ellipticity and position angle (PA), with cluster radius averaged across our sample, while Sect.~\ref{sec:average_shape_profiles_stacked_images} extends this analysis to stacked profiles in redshift bins. In Sect.~\ref{sec:shape_alignment_BCG_ICL_galaxies} we examine the alignment of the ICL with the BCG and large-scale satellite galaxy distribution of clusters. Finally, we summarise our findings in Sect.~\ref{sec:conclusion}.

Throughout this work, we adopt a flat $\Lambda$CDM cosmology with the Hubble parameter $H_{0}=70\,\mathrm{km\,s^{-1}\,Mpc^{-1}}$ and $\Omega_{\mathrm{m}}=0.3$. We use the convention that $M_{X\mathrm{c}}$ means the total mass enclosed within $R_{X\mathrm{c}}$, which is the radius inside which the density equals ${X}$ times the critical density of the Universe at the redshift of the cluster.

\section{Observations and methodology}
\label{sec:observations_sample}

\subsection{Cluster sample and properties}
\label{sec:cluster_sample}
The clusters used in this work were selected from two catalogues, each of which includes identified BCGs. We select the 76 clusters with halo masses of $M_{200\mathrm{c}} \geq 10^{14}\,M_\odot$ from the first eROSITA All-Sky Survey (eRASS1, \citealt{Bulbul_2024, Kluge_2024}) that overlap with the survey area of the Euclid Q1 data release. 
Spectroscopic and photometric properties (e.g. redshifts, optical centre, BCG position and velocity dispersions) of catalogue members, as well as the completeness and quality assessments of the survey, are discussed in \citet{Kluge_2024}.

We expanded our sample by including clusters from the Dark Energy Survey (DES) Y3 \citep{DES_25}, which has 174 clusters overlapping with the Q1 region. These clusters have $M_{200\mathrm{c}} \geq 10^{14}\,M_\odot$ and were constructed using photometric data from the redMaPPer\footnote{The redMaPPer catalogues can be accessed at \url{https://des.ncsa.illinois.edu/releases/y3a2/Y3key-cluster}.} catalogue \citep{Sevilla-Noarbe_2021}. Since the DES catalogue reports halo masses as $M_{200\mathrm{m}}$, we converted these to $M_{200\mathrm{c}}$ using the \texttt{Colossus} package \citep{Diemer_2018}, adopting an NFW density profile \citep{Navarro_1997} and the mass--concentration relation of \citet{Duffy_2008}. The same conversion was applied to the eRASS1 $M_{500\mathrm{c}}$ masses. Physical radii $R_{200\mathrm{c}}$ were then computed from $M_{200\mathrm{c}}$ and the critical density at the redshift of each cluster. 

To avoid duplicating clusters appearing in both catalogues, we cross-matched the BCG coordinates using a maximum on-sky separation of 10\,arcsec. This threshold is small enough to ensure we are matching the same cluster core, while allowing for cases where the two catalogues pick different BCG nuclei within the same system. The combination of the subsamples of eRASS1 and DES Y3 redMaPPer catalogues resulted in a final sample of 205 clusters for our analysis, spanning a halo mass range of $10^{14}\leq M_{200\mathrm{c}}/M_\odot\leq10^{15.2}$ and a redshift range of $0.1\leq z \leq0.8$.

To ensure accuracy and consistency of BCG selection across our sample and to correct for cases where the initial catalogue coordinates of the BCG were offset from their centre, we developed a custom routine using \texttt{photutils} \citep{Bradley_2025} and \texttt{astropy} \citep{astropy:2022} to refine the BCG coordinates based on image data. Using the initial coordinates, we constructed a cutout image of the cluster centred on the given location. 
After estimating and subtracting the background using a sigma-clipped median estimator, over a mesh of $10\times10$ pixel boxes tiled across the full image (with the background mesh smoothed using a $3\times3$ box filter), we applied the \texttt{photutils} source-detection algorithm \texttt{detect\_sources} to isolate bright sources via a segmentation map. We then deblended the overlapping sources and extracted source properties from the resulting catalogue. The BCG is selected as the object closest to the initial input coordinate in sky position. For further validation, we also locate the brightest pixel within the BCG segmentation mask. After evaluating the positions obtained from both the centroid and the brightest pixel, we retain the coordinate that best aligns with the BCG based on our visual inspection; in most cases corresponding to the brightest pixel. This step is necessary because the BCG centres reported in the original cluster catalogues were derived from low-resolution ground-based imaging, which occasionally blends multiple BCG nuclei or nearby companions that are readily resolved in the higher-resolution \Euclid\ images. The resulting offsets are typically only a few pixels.

\begin{figure}
    \centering
    \includegraphics[width=1\linewidth]{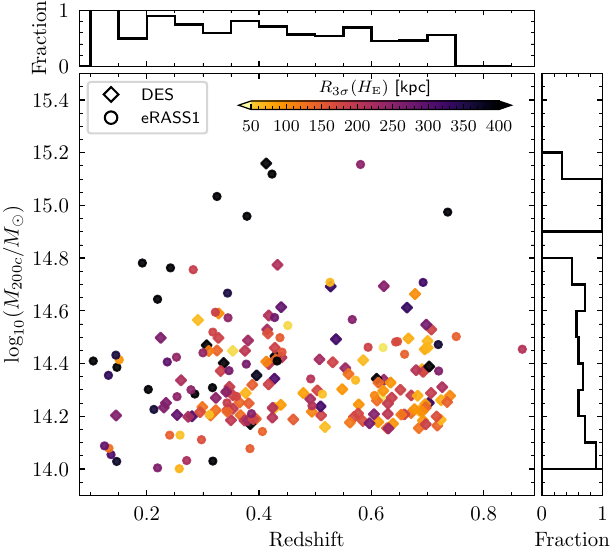}
    \caption{Cluster halo mass $M_{200\mathrm{c}}$ and redshift distribution of the cluster sample used in this work. Diamond and circular markers correspond to clusters from the DES and eRASS1 catalogues, respectively. The colour indicates the radius $R_{3\sigma}$ at which the ICL surface brightness profile in the \HE-band image reaches the 3$\sigma$ noise level (Sect.~\ref{sec:methods:Data_extraction_BCG_ICL_shapes}). Upper and right sub-panels respectively show for each redshift and mass bin the fraction of clusters for which we can measure the ellipticity of the ICL at $>$\,$0.1\,R_{200\mathrm{c}}$ relative to the total number of clusters in the same bin.}
    \label{fig:Sample_M200_redshift}
\end{figure}

The main panel of Fig.~\ref{fig:Sample_M200_redshift} shows the distribution of ${M_{200\mathrm{c}}}$ as a function of redshift for the cluster sample used in this study. The data points are colour-coded to show the semi-major axis extent to which we can observe the ICL surface brightness in the \HE images. Further details of these measurements are provided in Sect.~\ref{sec:methods:Data_extraction_BCG_ICL_shapes}.

\subsection{Q1 images of clusters}

The data analysed in this work originate from the publicly released Q1 \citep{Q1-TP001} data set. Our study focuses on the Euclid Deep Field South (EDF-S) and Fornax (EDF-F) fields, which overlap with our cluster sample. Observations were conducted with the 1.2-m space telescope of \Euclid in four photometric bands: the \IE filter in optical wavelengths using the VIS \citep{EuclidSkyVIS} instrument, and \YE, \JE, \HE filters in near-infrared (NIR) wavelengths using the NISP \citep{EuclidSkyNISP} instrument. 

To obtain low surface brightness preserved images of the clusters analysed in this work, we started from the Level 2 (LE2) single-exposure products generated by the Science Ground Segment (SGS) VIS \citep{Q1-TP002} and NIR \citep{Q1-TP003} pipelines. LE2 products consist of single-exposure, flat-fielded, astrometric- and flux-calibrated frames. We then processed them using the dedicated pipeline described in \citet[][hereafter B26]{Q1-Bamford}, which is optimised for studies of diffuse low surface brightness features around galaxy clusters.

The \citetalias{Q1-Bamford} processing includes the removal of residual atemporal background signals, correction of time-dependent instrumental artefacts, and subtraction and masking of charge persistence features in the NIR images. Although the native pixel scale of the VIS instrument is 0.1\,arcsec, such high resolution is not needed for ICL measurements, so we resample the images onto a common grid at the native resolution of NISP of $0.3\,\mathrm{arcsec\,pixel^{-1}}$. The images are subsequently background-matched across detector chips and coadded into a final $0\fdg5 \times 0\fdg5$ image for each cluster in the \IE, \HE, \JE, and \YE bands. In addition, a coadded $YJ\HE$ image is produced by combining the \HE, \JE, and \YE images weighted by their relative bandwidths. The overall aim of this processing is to suppress large-scale instrumental background structures while preserving faint extended emission. A detailed description of each pipeline step and their combined impact on ICL measurements is provided in \citetalias{Q1-Bamford}.

\subsection{Extraction of shape and surface brightness profiles of clusters}
\label{sec:methods:Data_extraction_BCG_ICL_shapes}

\begin{figure*}
    \centering
    \includegraphics[width=1\linewidth]{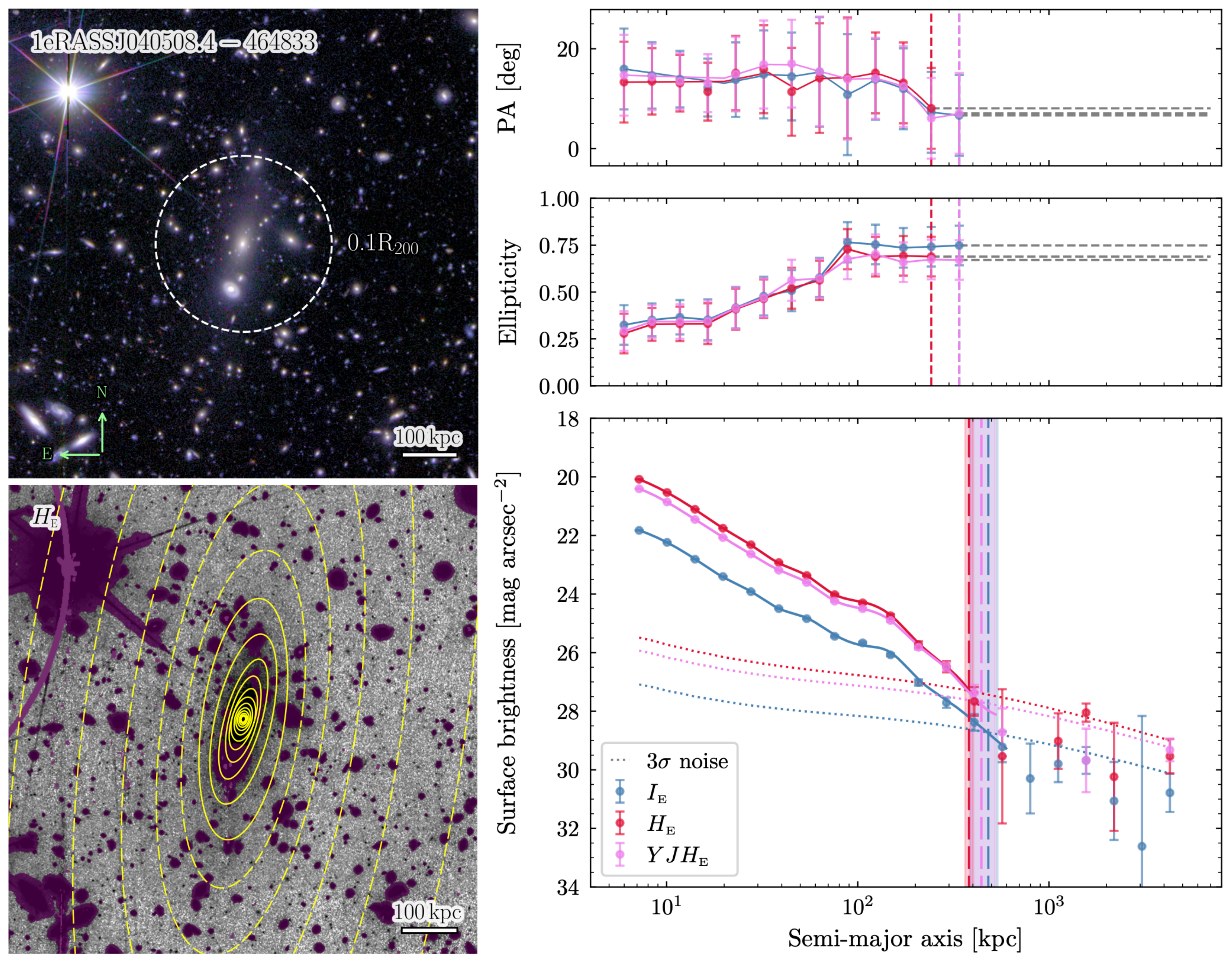}
    \caption{Images and extracted ICL profiles of a representative cluster (1eRASS J040508.4$-$464833) from our sample. \emph{Top left}: 
    Colour image constructed by combining all \Euclid bands. The white dashed circle marks the $0.1\,R_{200\mathrm{c}}$ ($\sim$175\,kpc) radius of the cluster. \emph{Bottom left}: \HE-band image with the source mask overplotted in purple. The yellow solid contours were constructed using \texttt{AutoProf} measurements from isophotal fitting, and yellow dashed contours take the exact shape of the last robust fit obtained at $R_{\rm shape}$.
    \emph{Top and middle right}: Radial PA profile of cluster light (measured east of north) and radial ellipticity profile, respectively, extracted from \IE, \HE, and $YJ\HE$ images. The dashed vertical lines represent the $R_{\rm shape}$ for each band. \emph{Bottom right}: Surface brightness profiles measured in elliptical annuli of the $YJ\HE$ image to ensure consistent comparison across bands. Interpolated spline fits are shown for each profile. Dotted lines represent the $3\sigma$ noise level, and vertical dashed lines mark the corresponding $R_{3\sigma}$ value for each band. Shaded regions, coloured to match each profile, indicate the uncertainty range of $R_{3\sigma}$.
    }
    
    \label{fig:Method_diagnostics}
\end{figure*}

Following the construction of the cluster images, we mask high surface brightness sources, estimate and subtract the large-scale background level, and finally measure the ICL surface brightness and shape using elliptical isophotes. All steps and procedures are described in detail in \citetalias{Q1-Bamford}, and we only summarise here the elements most relevant to this study.

We first construct a mask to identify the region dominated by the diffuse ICL emission using an adaptive smoothing algorithm. We smooth the image by a Gaussian of $\sigma_\mathrm{s}$ of up to 20 pixels and detect sources at a threshold of twice the background noise (measured in a box of 300\,pixels on a side). The mask covering the BCG+ICL is identified and then dilated so it encompasses the extent of the diffuse light in the cluster. For the clusters in our sample, this region typically extends over scales of approximately 100--150\,kpc. To detect the high surface brightness sources located within the BCG+ICL masked region, we first filter the image using a Gaussian median filter with a smoothing scale of $0.25\sigma_\mathrm{s}$. 
Then sources detected using a threshold of twice the noise above the local background are aggressively deblended. The detected sources (except for the BCG) are then grown by iteratively dilating until they are 25\% larger than their original size. 

To mask sources outside the BCG+ICL mask region, the image is smoothed with a Gaussian kernel ($\sigma = \qty{1}{pixel}$) and sources are detected above a threshold of three times the background noise (measured in a box of 500\,pixels) and requiring at least five detected pixels per source. The detected sources are then deblended using \texttt{photutils detect\_sources} and their individual masks are expanded by 50\% to minimise contamination from their extended haloes. 
For the NIR data, a single source mask derived from the coadded $YJ\HE$ image is applied consistently across all NIR bands, while the \IE-band image is masked independently. All of the choices for masking were determined and checked by testing on a set of mock images of simulated clusters. The results of those tests, shown in \citetalias{Q1-Bamford}, reveal that our approach provides adequate masking while avoiding over-masking of genuine ICL flux.

We then mask all sources and estimate a global background to the clusters. For each cluster, in each filter image, we estimate the background in boxes of 1\,Mpc in length (calculated at the redshift of the cluster) after the masks described above are applied to the images. We then filter the background over $3\times3$ grid of boxes to reduce sensitivity to any residual ICL flux that was not masked. This ensures that the background is estimated from a region at least 1\,Mpc from the BCG. After this background is subtracted, the BCG+ICL mask is removed and we extract shape profiles of the BCG+ICL using the elliptical isophote fitting tool \texttt{AutoProf} \citep{Stone_2021}. The images are centred and fixed on BCG positions. During the fitting, the background level parameter, \texttt{ap\_set\_background}, is set to zero to prevent an additional background subtraction that could lead to oversubtraction of the ICL. Further, a sequence of concentric elliptical isophotes is defined, with semi-major axes starting at 1\,pixel and increasing logarithmically by 0.15\,dex between successive isophotes, which increases the S/N in the outer regions where the ICL is faint.

In the fitting procedure, the centres and semi-major axes of the isophotes are held fixed, while \texttt{AutoProf} iteratively varies the ellipticity and PA of each isophote to best trace the underlying light distribution. The ellipticity is defined as $e = 1 - b/a$, where $a$ and $b$ are the semi-major and semi-minor axes, respectively. The isophote parameters are determined through Fourier analysis of the intensity distribution, with the second Fourier mode used to quantify departures from an ideal elliptical shape. To ensure stability at low surface brightness, the fitting incorporates a regularisation scheme that penalises large changes in ellipticity and PA between adjacent isophotes, resulting in smoothly varying shape profiles as a function of radius. The fitting is performed out to the semi-major axis where the local surface brightness per pixel reaches a signal-to-noise ratio of unity, as determined internally by \texttt{AutoProf}. We refer to the outermost semi-major axis at which the isophotal shape parameters can be robustly measured as $R_{\rm shape}$. Beyond this radius, the ellipticity and PA are held fixed at their last reliably determined values, while the isophotes are extended to larger radii for photometric measurements.

To ensure consistent measurements and enable direct comparison of the ICL surface brightness across wavebands, we measure the surface-brightness profiles in each band using the same isophotes derived from the coadded ($YJ\HE$) image. The surface brightness between each pair of adjacent isophotes is calculated as the median pixel value within the corresponding elliptical annulus of the background-subtracted image, after masking high surface brightness sources.

In our measurements, we adopt the limiting surface brightness as the faintest surface-brightness level that can be securely detected above the noise at a given confidence level. This limit is determined by characterising the photometric noise in the post-processed images arising from residual background inhomogeneities, and is evaluated separately for the \IE, \YE, \JE, \HE, and $YJ\HE$ bands. The surface-brightness profiles of the noise used in this work are adopted from \citetalias{Q1-Bamford}, and are derived by measuring surface-brightness fluctuations within elliptical apertures evaluated at approximately 1000 randomly selected positions across large $\ang{2.5}\times\ang{2.5}$ regions of each Q1 survey field, following source masking and subtraction of large-scale background variations. These diagnostics show that on the scale of several arcminutes, we can detect surface brightness to $3\sigma$ level, around $27.5\,\mathrm{mag\,arcsec^{-2}}$ in \YE, \JE, \HE, and $29\,\mathrm{mag\,arcsec^{-2}}$ in \IE.

Uncertainties on the BCG+ICL surface-brightness profiles are estimated as the scatter in the resulting surface-brightness values in each annulus defined via the median absolute deviation, rescaled to a $1\sigma$ uncertainty. Background-related uncertainties dominate at large radii, while Poisson noise is significant only in the central BCG regions; these contributions are combined in quadrature to obtain the total uncertainty.

To deduce the extent of ICL measurements we apply a weighted spline fitting to the surface brightness profiles (measured using the ellipses defined in the $YJ\HE$ coadded images). We use a logarithmic radius scale to capture the dynamic range of profiles and use the inverse variance of the surface brightness noise as the weights to emphasise points with small and large uncertainties. Using \texttt{UnivariateSpline} procedure of \texttt{scipy} Python package \citep{Virtanen_2020SciPy-NMeth}, a cubic ($k=3$) interpolated spline is then fitted to valid data points above the $3\sigma$ noise level, since measurements fainter than this threshold show large fluctuations and are unreliable. We quantify the extent of the ICL as the semi-major axis at which the fitted spline crosses the surface brightness of $3\sigma$ noise curve -- defining this semi-major axis as $R_{3\sigma}$. 

To estimate the uncertainty in $R_{3\sigma}$, we perform bootstrap resampling by perturbing the fluxes within their noise distributions over 1000 iterations. For each iteration, an interpolated spline is fitted and the corresponding $R_{3\sigma}$ value is extracted. The median of the resulting $R_{3\sigma}$ distribution is adopted as the final measurement, while the 16th and 84th percentiles define the lower and upper uncertainties, respectively.

Figure~\ref{fig:Method_diagnostics} shows example images and profiles for a representative cluster (1eRASS J040508.4$-$464833) in our sample, at $z=0.37$ with a halo mass of $8.8\times10^{14}\,M_\odot$ ($R_{200\mathrm{c}}\approx1490\,\mathrm{kpc}$). In the right-hand panels, the top and middle panels show the PA and ellipticity profiles as a function of semi-major axis, measured independently in each band using \texttt{AutoProf}; the vertical dashed lines indicate the corresponding $R_{\rm shape}$ values. Uncertainties on both the PA and ellipticity are derived by analysing a set of mock cluster observations and exploring how accurately we can measure the true PA and ellipticity. We derive the scatter between the true PA and ellipticity of the mock clusters and the PA and ellipticity measured via \texttt{AutoProf}. The plotted uncertainties in Fig.~\ref{fig:Method_diagnostics} are the median absolute deviation, rescaled to a $1\sigma$ uncertainty. Full details of how we measure the reliability of our PA and ellipticity measurements can be found in \citetalias{Q1-Bamford}.

The bottom-right panel shows the surface-brightness profiles, measured on a consistent set of elliptical isophotes defined from the $YJ\HE$ coadded image to enable direct comparison between bands. The vertical dashed lines and shaded regions mark the locations of the $R_{3\sigma}$ radii and their associated uncertainties. For the cluster shown in the figure, the $R_{3\sigma}$ values are consistent across bands, ranging from roughly 350 to 500\,kpc, corresponding to $\sim 27.5\,\mathrm{mag\,arcsec^{-2}}$ for \HE and $YJ\HE$ and $\sim 28.7\,\mathrm{mag\,arcsec^{-2}}$ for the \IE band. On the \HE-band image, shown in the bottom-left panel of the figure, the solid ellipses mark the isophotes from which a shape could be measured, while the dashed ellipses indicate the region where the shapes of the ellipses are fixed. The first dashed ellipse marks $R_{\rm shape}$. 

Throughout this work we characterise the shape of the ICL (defined as its PA and ellipticity) at a semi-major axis of $0.1\,R_{200\mathrm{c}}$, which corresponds to a typical length of 100\,kpc for most clusters within our sample. Adopting $0.1\,R_{200\mathrm{c}}$ as a representative scale for the ICL shape allows for characterisation at a radius where its surface brightness remains detectable while sufficiently separated from the central region where the BCG dominates the surface brightness profile (e.g. \citealt{Contini_2014}). 

We note that the clusters with $R_{\rm shape}$\,$<0.1\,R_{200\mathrm{c}}$ are excluded from some parts of our analysis because in these clusters we are able to measure the shape of the ICL only where it overlaps with the BCG, which may introduce potential inconsistencies. Where this criterion is applied, it is explicitly stated. In practice, this criterion excludes 75 clusters, spanning the full redshift and halo mass range of our sample, leaving us with a sample of 130 clusters. The top and right sub-panels of Fig.~\ref{fig:Sample_M200_redshift} show the fraction of clusters with $R_{\rm shape}>0.1\,R_{200\mathrm{c}}$ in redshift and mass bins, respectively. Across the full mass range and at low to intermediate redshifts, the sample is at least $\sim$\,80\% complete, but this fraction decreases to $\sim$\,60\% at the highest redshifts ($z\gtrsim0.6$), where shape measurements become more challenging. This shows that the criterion does not preferentially remove low-mass systems with small $R_{200\mathrm{c}}$, but instead reflects limitations in the measured $R_{\rm shape}$ across the full range of halo sizes. 

\section{Sensitivity of \Euclid to ICL}
\label{sec:Euclid_filter_sensitivity_to_ICL}

\begin{figure}
    \centering
    \includegraphics[width=1\linewidth]{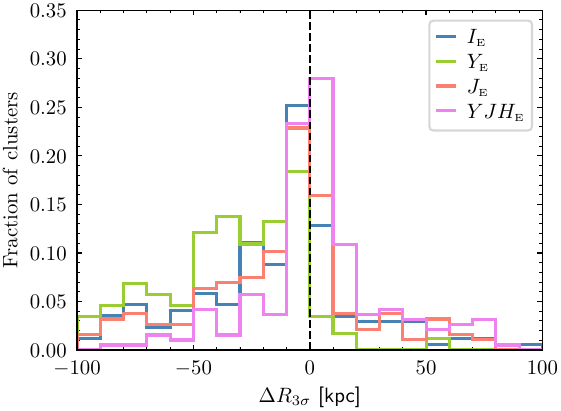}
    \caption{Difference between $R_{3\sigma}$ measured across photometric bands and in the \HE band, with colours indicating the difference measured in each band relative to \HE: \IE (blue), \YE (green), \JE (orange), and $YJ\HE$ (pink). The vertical dashed line marks the zero for guidance.}
    \label{fig:R3sigma_differences_hist}
\end{figure}

We first explore which of the \Euclid filters best supports the detection of ICL and the measurement of its shape properties. In Fig.~\ref{fig:R3sigma_differences_hist}, we present the difference between the $R_{3\sigma}$ measured in each of the four bands relative to the \HE band, hereafter being referred to as $\Delta R_{3\sigma}$. The distributions for $\Delta R_{3\sigma}(\IE,\HE)$ and $\Delta R_{3\sigma}(\JE,\HE)$ peak near zero or shift slightly towards negative values, approximately with mean difference of $\Delta R_{3\sigma}(\IE,\HE)=-7.9$\,kpc and $\Delta R_{3\sigma}(\JE,\HE)=-6.3$\,kpc. This suggests that, on average, ICL can be detected to only slightly larger radii in the \HE band than in the \IE and \JE bands. The $\Delta R_{3\sigma}(\YE,\HE)$ distribution is the broadest and is skewed towards negative values with a mean $\Delta R_{3\sigma}(\YE,\HE)=-30$\,kpc. This indicates that the \YE band is the least sensitive of the \Euclid filters for detecting the extended ICL. The sensitivity is limited at bluer wavelengths because the ICL emission is intrinsically fainter, while at NIR wavelengths it is further reduced by the increasing contribution of zodiacal light around $\sim$\,1\,$\mu$m. Since \YE has the same exposure time as \HE and \JE, these factors together make the ICL signal weakest in the \YE band. Although the \IE band probes even shorter wavelengths than \YE, it benefits from the superior sensitivity and greater exposure time. These advantages help mitigate the effects of lower ICL flux in the optical regime, enabling competitive ICL detection compared to the \HE band. In contrast, the $\Delta R_{3\sigma}(YJH_{\rm E},\HE)$ distribution is slightly skewed towards positive values with a mean difference of $1.8$\,kpc. This is consistent with $YJ\HE$ images having slightly greater sensitivity to the ICL compared to the \HE band. However, this improvement is small compared to the width of the distribution.

\begin{figure}
    \centering
    \includegraphics[width=1\linewidth]{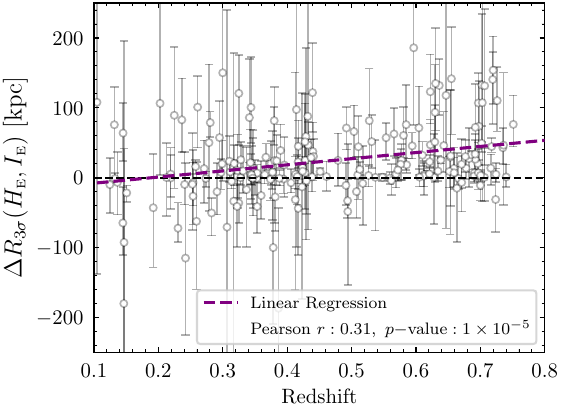}\hfill
    \caption{Difference in radius $R_{3\sigma}$, where the ICL surface brightness profile reaches $3\sigma$ noise level, measured in the \HE and \IE bands as a function of cluster redshift. The dashed purple line shows the best-fitting linear regression.}
    \label{fig:R3sigma_BEST_Z}
\end{figure}

\begin{figure}
    \centering
    \includegraphics[width=1\linewidth]{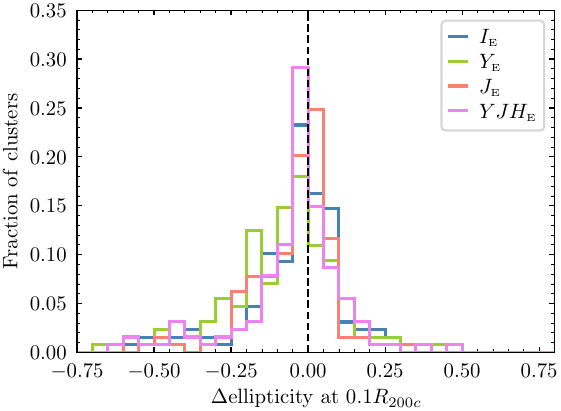} 
    \caption{Distributions of ICL ellipticity differences between the \HE band and other photometric bands. The colours represent the distribution obtained for each band (same as in Fig.~\ref{fig:R3sigma_differences_hist}).}
    \label{fig:Ellipticity_diff_across_bands_0.1R200}
\end{figure}

\citet{Q1-SP008} show that the ICL has a red colour in the \Euclid bands and becomes progressively redder with increasing redshift. Motivated by this, we assessed whether the \IE band provides a better measure of ICL at low redshift compared to the \HE images. In Fig.~\ref{fig:R3sigma_BEST_Z}, we show the difference between the $R_{3\sigma}$ measurements in the \HE and \IE bands as a function of redshift. The error bars for each point were derived by propagating the uncertainties of the two independent $R_{3\sigma}$ measurements used to calculate $\Delta R_{3\sigma}$ between the \HE and \IE bands for each cluster. We find a weak but statistically significant positive Pearson correlation ($r = 0.31$, $p = 1\times10^{-5}$), indicating that the relative extent measured in \HE increases with redshift. Estimating the zero-crossing redshift of the best-fitting relation via bootstrap resampling yields a median value of $z=0.2$, with a 16--84\% range of $0.07$--$0.26$. However, this transition is constrained by a relatively small number of clusters at $z<0.2$ in our sample, which contributes to the substantial intrinsic scatter in $\Delta R_{3\sigma}$ and limits the precision with which the crossing redshift can be determined. Below this redshift, the measurements are nevertheless consistent with the \IE band providing a comparable, and in some cases more complete, recovery of the ICL. In summary, it is advantageous to use the \HE band to measure the ICL at higher redshifts, while the \IE band remains well suited for ICL studies at $z<0.2$.

Since we are primarily concerned with the shape measurements of ICL, we furthermore explore how the ICL ellipticity measurements vary across different photometric bands. Figure~\ref{fig:Ellipticity_diff_across_bands_0.1R200} shows the differences in ICL ellipticities, measured at $0.1\,R_{200\mathrm{c}}$, across the \IE, \YE, \JE, and $YJ\HE$ bands relative to the \HE band. Clusters with $R_{\rm shape}<0.1\,R_{200\mathrm{c}}$ are excluded. The distributions exhibit negligible systematic offsets relative to the \HE-band distribution, with median ellipticity differences ranging approximately from $-0.02$ to $-0.05$. For each band, $\sim$\,90\% of clusters have ellipticity differences within $\pm$0.3 compared to \HE. In examining the remaining $\sim$\,10\% of clusters that exhibit large ellipticity variations across bands, we find that the differences are driven by residual systematic errors, including imperfect masking of foreground stars, stellar diffraction spikes, or the extended haloes of bright member galaxies, which influence the shape measurements differently in each band. Taken together, these results demonstrate that shape measurements are broadly consistent and robust across photometric bands in 90\% of the clusters. We therefore limited the remaining analysis of shape measurements to the \HE band as the reddest band in our study, which also on average traces the ICL to the furthest radii (Fig.~\ref{fig:R3sigma_differences_hist}). An alternative approach would be to restrict our comparison to these 90\% of clusters, excluding systems with the largest inter-band differences. While this would reduce the scatter, it risks introducing selection effects by removing those clusters where the ICL is faintest or most irregular. We therefore retained the full sample, noting that the small fraction of outliers does not alter our conclusion that ellipticity measurements are robust across bands.

The $R_{3\sigma}$ values measured in the \HE band across our sample are shown as the colour scale in Fig.~\ref{fig:Sample_M200_redshift}, spanning a broad range from $\sim$\,25\,kpc to values exceeding 400\,kpc. Using Pearson's correlation coefficients, we find a weak negative correlation with redshift ($r=-0.27$, $p=1.2\times10^{-4}$) and a weak positive correlation with halo mass ($r=0.30$, $p=1.2\times10^{-5}$). Because the relations may not be strictly linear, we also compute Spearman's rank coefficients, which yield $\rho=-0.28$ ($p=4.3\times10^{-5}$) for redshift, and $\rho=0.23$ ($p=7.5\times10^{-4}$) for halo mass, confirming the same trends. These results indicate that while more massive haloes host ICL detectable to larger radii, the observed suppression of $R_{3\sigma}$ at higher redshifts is likely driven by redshift-dependent effects such as cosmological dimming.

We observe consistent trends in the ICL extent across redshift--mass space. At $z\sim0.2$ and $\log_{10}(M_{200\mathrm{c}}/M_\odot) \sim 14.2$, $R_{3\sigma}$ values typically range between 150 and 250\,kpc. For slightly more massive clusters at $z \sim 0.3$--$0.4$ and $\log_{10}(M_{200\mathrm{c}}/M_\odot) \sim 14.5$, the extent can increase to 300--400\,kpc or beyond. At higher redshifts $z \sim 0.7$, clusters with similar halo masses exhibit lower $R_{3\sigma}$ values of less than 150\,kpc. These trends are consistent with the model predictions presented in figure~12 of \citet{EP-Bellhouse}.

\begin{figure}
    \centering
    \includegraphics[width=1\linewidth]{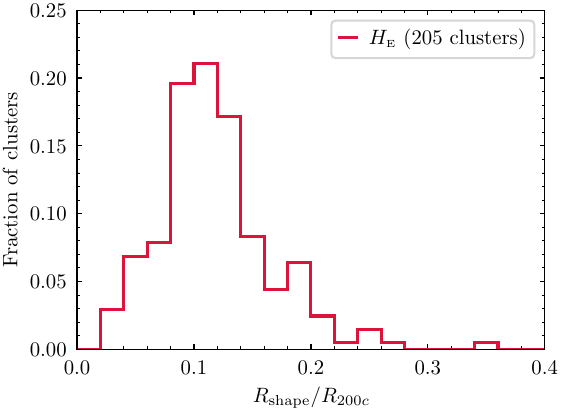}
    \caption{Radial extent of the ICL shape measurement of clusters in our sample in the \HE band relative to their $R_{200\mathrm{c}}$.}
    \label{fig:rshape_over_r200}
\end{figure}

In Fig.~\ref{fig:rshape_over_r200} we show how far out we can measure the shape of the ICL ($R_{\rm shape}/R_{200\mathrm{c}}$) in the \HE images for each cluster in our sample. We can measure the shape of the ICL out to $0.1\,R_{200\mathrm{c}}$ for $63\%$ of our sample, while $\sim$6\% of the clusters have shape measurements beyond $0.2\,R_{200\mathrm{c}}$. These results demonstrate that the shape of the diffuse light distribution in individual clusters can typically be traced out to $\sim$\,10\% of $R_{200\mathrm{c}}$ in our data, corresponding to a physical extent of $\sim$\,100\,kpc for the typical halo mass of our cluster sample.

\section{Distribution of ICL ellipticity}
\label{sec:BCG_ICL_ellipticity_distribution}

\begin{figure}
\centering
    \includegraphics[width=1\linewidth]{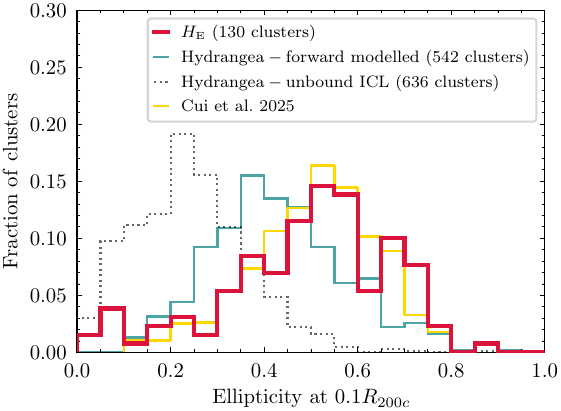} 
    \caption{Distribution of ICL ellipticities measured at $0.1\,R_{200\mathrm{c}}$. Distributions are shown in red for \HE-band observations, grey (dotted) for Hydrangea mocks with unbound ICL only, and blue for forward-modelled mocks injected into a realistic \Euclid sky background. The yellow distribution shows cluster ellipticities derived from gravitational lensing measurements from \citet{Cui_25}.}
    \label{fig:Ellipticity_at_ICL_Hydrangea_H}
\end{figure}

The distribution of ICL ellipticity, measured at $0.1\,R_{200\mathrm{c}}$, is presented in Fig.~\ref{fig:Ellipticity_at_ICL_Hydrangea_H}. The distribution peaks at a median ellipticity of $0.513$ (mean $e=0.489\pm0.01$), and exhibits a broad standard deviation ($\sigma\simeq0.179$). Approximately 9\% of clusters have much rounder ICL haloes than the median, with $e\le0.2$, while 26\% have significantly more elliptical haloes, with $e\ge0.6$. We note that, upon examining the most elliptical haloes in our sample ($e\geq 0.6$; 34 clusters in total), we find that the ellipticity measurements of 15 clusters are affected by residual systematic errors, including imperfect masking of foreground stars, contamination from member galaxies, or light from the extended halo of a bright member galaxy, which can artificially lead to more elongated halo shapes. However, most of our highly elliptical clusters exhibit high intrinsic halo ellipticities that cannot be explained by residual systematic effects in the data. 

Our mean ICL ellipticity measurement is in excellent agreement with direct constraints on dark matter halo shapes from gravitational lensing studies of clusters. Using a strong-lensing sample of 39 massive clusters with virial masses of $M_{\rm vir}\simeq10^{14}$--$10^{15}\,M_\odot$ over the redshift range $0.2\lesssim z \lesssim 0.9$, \citet{Okabe_2020b} reported a mean projected halo ellipticity of $0.482\pm0.028$, in agreement with our mean ICL ellipticity. Similarly, a weak lensing analysis of 25 X-ray-selected clusters with comparable virial masses ($M_{\rm vir}\simeq10^{14}$--$10^{15}\,M_\odot$) at $0.15<z<0.3$ by \citet{Oguri_2010} found a mean halo ellipticity of $0.46\pm0.04$, again consistent with our result.

As a further comparison, in Fig.~\ref{fig:Ellipticity_at_ICL_Hydrangea_H}, we overlay the distribution of halo ellipticities measured from weak lensing shear maps by \citet{Cui_25}, as they have similar redshift range and sample size as our cluster sample. Their analysis, based on 299 redMaPPer-selected clusters from the HSC-SSP survey at $0.08<z<0.55$, fitted elliptical NFW models directly to two-dimensional shear maps. The resulting distribution has a median ellipticity of $\simeq$\,0.52 and a scatter of $\simeq$\,0.13, closely resembling our \HE-band ICL ellipticity distribution: both peak around ellipticities of $\sim$\,0.5 and extend to highly elongated systems with values $\gtrsim$\,0.6. Although the scatter is somewhat larger in the ICL distribution, the overlapping tails towards both rounder and more elongated shapes suggest that ICL and lensing capture a comparable range of intrinsic cluster morphologies. We note that, analogous to our ICL measurements, weak lensing estimates of halo shapes may also be affected by line-of-sight projections and observational limitations, which can bias measurements towards more elongated values. Cluster selection, as well as cluster mass and dynamical state, can influence the inferred ICL shape measurements. However, the limited size of the current sample restricts our ability to explore these effects in detail. We plan to investigate this dependence more thoroughly using the much larger cluster sample expected from Euclid DR1, which will comprise several thousand systems.

Despite methodological differences, especially that our ICL shapes are extracted at a fixed fractional radius while \citet{Cui_25} and other lensing studies infer global projected halo ellipticities at a different radial regime, the convergence of medians across these independent probes demonstrates a key result. Since lensing directly traces the dark matter halo, the agreement with our ICL measurements suggests that the ICL is indeed measuring the shape of the dark matter halo itself. Taken together, these results establish a robust ellipticity range of $\sim$\,0.45--0.55 for cluster haloes, consistent with earlier studies \citep[e.g.][]{Ho_2006,Evans_2009,Oguri_2010,Despali_2017,Okabe_2020b,Cui_25,Liu_2025}. 

Moreover, \citet{Gonzalez_2022} used weak lensing stacking on cluster-mass haloes in the MICE-GC simulation and expressed projected shapes via $q=b/a$. Their dark matter particle shapes peak around $q\sim0.6$--0.7 ($e\approx 0.3$--0.4), overlapping the moderate-ellipticity regime of our ICL distribution, and they show that miscentring/misalignment and dynamical complexity can bias stacked lensing shape estimates, consistent with our interpretation of the high-$e$ tail seen in the observed distribution. The close agreement between the ICL ellipticities and those inferred for dark matter haloes strongly supports ICL as a reliable luminous tracer of the underlying cluster potential across cosmic time. 

To test whether our results agree with theoretical expectations, we compared the ICL shapes inferred from observations with those derived from mock clusters generated by the Hydrangea cosmological hydrodynamic simulations \citep{Bahe_2017}, which is a suite of zoom-in cluster simulations that uses the EAGLE galaxy formation model \citep{Schaye_2015}. The simulated clusters were selected at redshifts similar to those of the clusters analysed in this work ($z = 0.2,\ 0.4,\ 0.5,$ and $0.7$) from a gigaparsec-scale parent simulation \citep{Barnes_2017_CLEAGLE} and therefore cover the full mass range of our \Euclid sample ($M_{200\mathrm{c}}\sim10^{14}$--$10^{15}\,M_{\odot}$).

We generated \Euclid-like mock images of the Hydrangea clusters in the $xy$, $xz$ and $yz$ projections, following the approach of \citet{martin_preparing_2022}, yielding a sample of 636 mock clusters. A pixel scale of $0.3\,\rm{arcsec\,pixel^{-1}}$ is adopted to match the resolution of \Euclid. The stellar particles are extracted in three dimensions, and their spectral energy distributions were assigned according to their individual ages and metallicities. We extracted all stellar particles that are located within $R_{200\mathrm{c}}$ of the most bound particle. Each spectral energy distribution was then redshifted to the cluster frame, passed through the \Euclid \HE-band transmission curve and corrected for stellar mass loss to obtain apparent magnitudes. To reduce pixel-scale artefacts in diffuse regions, the particle distribution was smoothed prior to projection. Finally, the images were convolved with the extended \HE-band \Euclid point-spread function (PSF; \citealt{EROData}).

These mock clusters were injected into randomly selected regions of a large \Euclid sky patch (Sect.~\ref{sec:methods:Data_extraction_BCG_ICL_shapes}). This procedure placed the intrinsically noiseless clusters into real \Euclid backgrounds containing spatially varying noise, artefacts, and sky gradients. 
To measure the PA and ellipticity of the ICL, the mock images were analysed following the procedures applied to real \Euclid clusters as described in Sect.~\ref{sec:methods:Data_extraction_BCG_ICL_shapes} -- including masking, background subtraction, and shape extraction using \texttt{AutoProf}. These measurements result in a fully forward-modelled sample of 542 (85\% of the mock cluster sample) clusters with robust ellipticity measurements at $0.1\,R_{200\mathrm{c}}$.

The resultant distribution of ICL ellipticities from the mock images is shown in Fig.~\ref{fig:Ellipticity_at_ICL_Hydrangea_H} (labelled as Hydrangea -- forward modelled). We first note that ellipticity measurements at $0.1\,R_{200\mathrm{c}}$ could be achieved in 85\% of the mock clusters compared to 63\% of the real clusters observed by \Euclid. This is a result of the ICL in the simulations being brighter on average than it is in the real Universe \citep{Q1-Gully}. While we see a broad agreement, the simulated ICL is generally rounder than in the observations: the mean ellipticities of the distributions are $e=0.49\pm0.01$ for the observed clusters versus $e=0.43\pm0.01$ for the forward-modelled clusters. A Kolmogorov--Smirnov (KS) test comparing these distributions results in $D=0.27$ and $p=2.9\times10^{-7}$ indicating a statistically significant difference between the two samples. Given the limited halo diversity in the simulations\footnote{Our simulated cluster sample is drawn from a set of only 86 unique Hydrangea clusters that are placed at multiple redshifts and viewed along different projections. As a result, the effective diversity of halo shapes in the simulations is limited.} and the higher average ICL surface brightness in the forward-modelled clusters, we caution that this difference may reflect sample limitations rather than a physical disagreement between observed and simulated ICL morphologies. Nevertheless, different simulations result in different ICL morphologies due to different initial conditions and galaxy formation physics \citep{Puchwein_2010,Rudick_2011,Kluge_2021,Brough_2024,Fernandez_26}, so the disagreement seen in Fig.~\ref{fig:Ellipticity_at_ICL_Hydrangea_H} suggests that ICL halo shapes may be a useful constraint for large-scale simulations.

In simulations ICL is often defined as the light emitted from stars that are gravitationally bound to the cluster potential rather than any cluster satellite galaxy. We therefore created images of this unbound ICL component by excluding all stellar particles bound to satellite substructures that are identified using \textsc{Subfind} \citep{Dolag_2010}, and applying the PSF, but we do not place them into randomly selected regions of the \Euclid images. Thus, these unbound ICL images can be considered noiseless. We measure the ICL properties in these noiseless unbound-ICL images in the same way as described above, and plot their ellipticity distribution in Fig.~\ref{fig:Ellipticity_at_ICL_Hydrangea_H} (labelled as unbound ICL). Since there is no noise, we are able to measure the ICL properties in all 636 of the simulated unbound-ICL images.

Our method of isolating ICL in the \Euclid images attempts to capture the light from these unbound stars by masking satellite galaxies and their outskirts. In doing so, we mask regions where there is unbound ICL, but also leave in light from stars in the far outskirts of galaxies that are weakly bound to the satellite rather than the cluster. Cluster satellites are constantly being stripped of their stars so their outskirts are abundant with stars that are in the process of becoming ICL, but are not yet unbound. 

As shown in Fig.~\ref{fig:Ellipticity_at_ICL_Hydrangea_H}, the ellipticity of the unbound ICL is significantly rounder than the ICL that we measure via the masking of satellites, which highlights that full forward-modelling of the ICL is the only correct way to compare observed ICL with that in simulations. \citet{Fernandez_26} have shown that the unbound ICL is a reliable tracer of the shape of the underlying dark matter halo of clusters in simulations. Since our measured ICL is biased to more elliptical shapes than the unbound ICL, with mean ICL ellipticity of 0.43 for the forward-modelled sample against 0.24 for the unbound ICL sample, it is likely that our measured distribution of ICL is biased to be more elliptical than the underlying dark matter. With 2D shear maps provided by the Euclid DR1 release, we plan to empirically calibrate the relationship between the shape of the dark matter halo of a cluster and its measured ICL shape.

\section{Variation of ICL halo shape with cluster radius}
\label{sec:average_shape_profiles}

\begin{figure}
    \centering
    \includegraphics[width=1\linewidth]{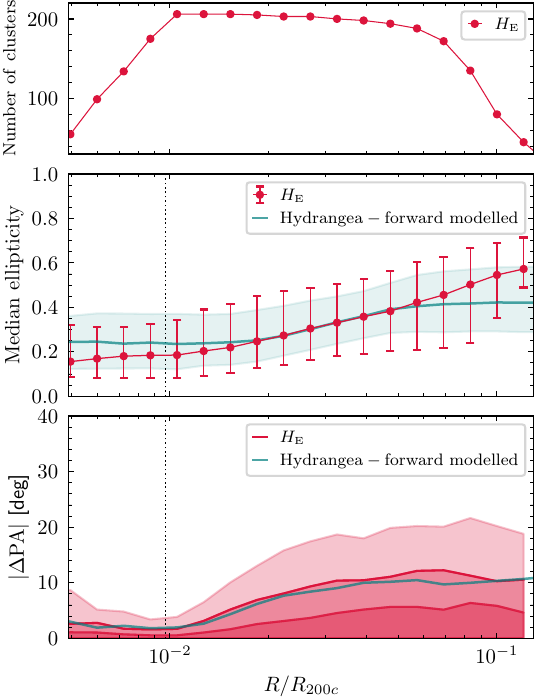} 
    \caption{Radial shape profiles of BCG+ICL averaged across our sample. The number of clusters contributing to the average measurements in each radial bin (top), median ellipticity (middle), and median PA difference between cluster profiles and the BCG measured at 10\,kpc (bottom) of clusters is shown as a function of $R/R_{200\mathrm{c}}$ across \HE band. Error bars in the middle panel indicate the 16th and 84th percentiles, while the shaded regions in the bottom panel represent the 25th, 50th, and 75th percentiles. Blue lines shown in the middle and bottom panels are the distributions obtained from forward-modelled Hydrangea mock cluster images. The vertical dotted lines mark the median $R_{200\mathrm{c}}$ position of the sample that corresponds to 10\,kpc. In both panels, the area left of this line represents the BCG region. }
    \label{fig:Average_ellipticity_PA}
\end{figure}

We characterised how the BCG+ICL shape varies as a function of radius in Fig.~\ref{fig:Average_ellipticity_PA}, where we have computed the median ellipticity and PA of our cluster sample within common radial bins. The number of clusters contributing to each radial bin is shown in the top panel. To ensure statistical reliability, we only included radial bins that contain measurements of at least 30 clusters (15\% of the total sample). The maximum radius at which this threshold is met lies in the bin just greater than $0.1\,R_{200\mathrm{c}}$, where $\sim$60 clusters contribute. We further excluded measurements that correspond to radial extents of less than two times the PSF FWHM, which accounts for the drop in the number of contributing clusters within the inner annuli.

We find a gradual increase in ellipticity with radius: in the inner regions, where the BCG dominates the light distribution, shapes are relatively round, with ellipticities just below 0.2. To indicate the BCG-dominated region in the distributions shown in Fig.~\ref{fig:Average_ellipticity_PA}, we adopt a characteristic physical BCG size of 10\,kpc \citep{Kluge_2020,Contini_2024_vegas} and convert this scale into the normalised radial coordinate $R/R_{200\mathrm{c}}$ for each cluster; the median of the resulting distribution is shown as a vertical line in the middle and bottom panels. At larger radii, the profiles become increasingly elongated, reaching ellipticities above 0.5, indicating that, on average, the ICL is more elliptical than the BCG.

A consistent picture across photometric and lensing studies is that the BCG-dominated inner light is relatively round, while the diffuse ICL becomes increasingly flattened with radius, yielding an outward rise in mean ellipticity. \Euclid ERO analyses show a clear transition beyond which BCG+ICL isophotes grow more elliptical, and earlier deep imaging similarly required multi-component models with independent ellipticities to reproduce observed ellipticity gradients \citep{EROPerseusICL,Gonzalez_2005}. Weak lensing halo-shape measurements likewise find larger projected dark matter ellipticities at larger radii, reinforcing the same qualitative trend from an independent probe of the cluster potential \citep[e.g.][]{Liu_2025}.

For comparison, in Fig.~\ref{fig:Average_ellipticity_PA} (middle panel), we include the median ellipticity distributions obtained from the forward-modelled Hydrangea mock clusters. Whereas the ellipticity at $0.05\,R_{200\mathrm{c}}$ is similar for simulations and the observed \Euclid clusters, we see that the simulated clusters are more elliptical in the BCG by $\sim 0.05$, and rounder at larger radii. While both the simulated and observed profiles display rounder BCGs and more elongated ICLs, the transition between these regimes is more gradual in the simulations. \citet{Fernandez_26} explore how the ellipticity of the BCG and ICL varies as a function of cluster radius and find that the shape of the BCG and ICL is highly dependent on the underlying physics of the simulations. In particular, the Hydrangea simulations exhibit more elliptical BCGs than other simulations, such as Horizon-AGN. Our measurements of the radial variations in BCG+ICL ellipticity can therefore be used as a benchmark to constrain the physics involved in the formation of BCGs and the ICL \citep{Fernandez_26}. 

The bottom panel of Fig.~\ref{fig:Average_ellipticity_PA} shows the median profiles of the PA offset of the clusters from their BCG PA values measured at 10\,kpc. This captures how the orientation of the light distribution within a cluster changes with radius compared to the BCG. We show the median (50th percentile) of these offsets, along with the 25th and 75th percentiles, to illustrate the variation across the sample. Within the BCG-dominated region ($R\lesssim0.01\,R_{200\mathrm{c}}$) the PA offset is small: the median distribution is around $5^\circ$ and the 75th percentile is below $10^\circ$. The median offset steadily increases with radius, reaching values of $\sim$\,10$^\circ$ near $0.1\,R_{200\mathrm{c}}$. For 75\% of clusters the PA varies by less than $\sim$\,20$^\circ$. Together with the significant ellipticity gradient this implies that the primary change in shape of the BCG+ICL to larger radii is a gradual stretch, with only a minor twist.

In the same panel, for comparison, we also show the median PA differences ($50^{\rm th}$ percentile) obtained from the forward-modelled Hydrangea mock clusters as the blue line. This profile reproduces the trend of increasing misalignment with radius well; following the median observational PA offsets across the radii. Both observations and simulations indicate that the BCG and ICL remain well aligned within $0.1\,R_{200\mathrm{c}}$, with typical PA differences of only $\sim$\,10$^{\circ}$.

\section{ICL halo shape in stacked clusters}
\label{sec:average_shape_profiles_stacked_images}
\begin{figure*}
    \centering
        \includegraphics[width=1\linewidth]{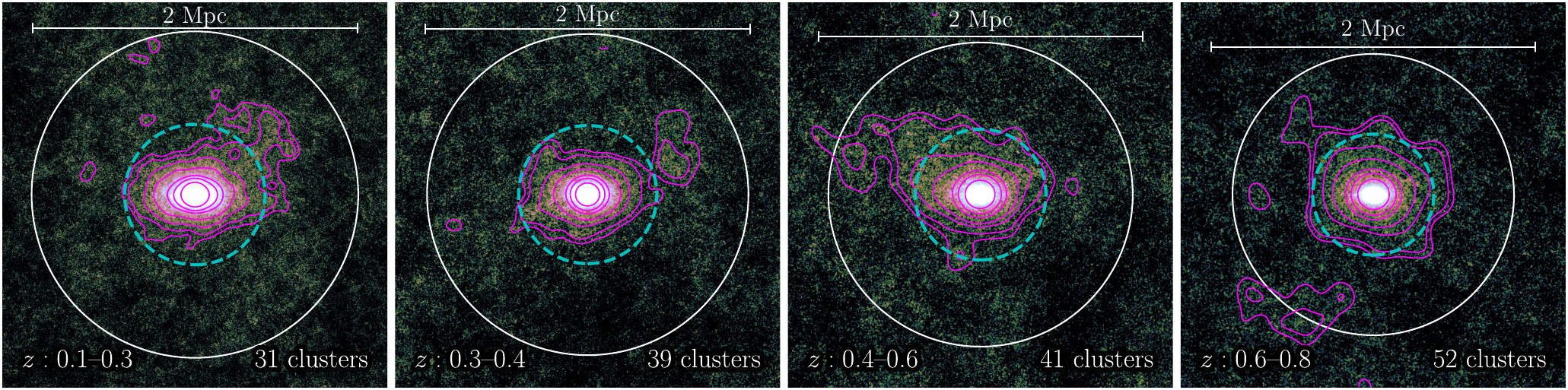}
    \caption{Stacked \HE-band images of clusters in different redshift bins. Prior to stacking, the member galaxies and bright sources were masked, and the position angles of the clusters were aligned to the $x$ axis (as we explain in Sects.~\ref{sec:methods:Data_extraction_BCG_ICL_shapes} and \ref{sec:average_shape_profiles_stacked_images}). The white circle in each panel indicates the $R_{200\mathrm{c}}$ of the stacked cluster, while the dashed cyan circle marks $0.4\,R_{200\mathrm{c}}$. The magenta contours trace the surface brightness distribution at the 85th, 90th, 95th, 97th, 98.5th, 99th, and 99.5th percentiles of the flux. A 2\,Mpc scale bar is shown for reference.}
    \label{fig:H_stacked_images}
\end{figure*}

\begin{figure*}
    \includegraphics[width=0.5\linewidth]{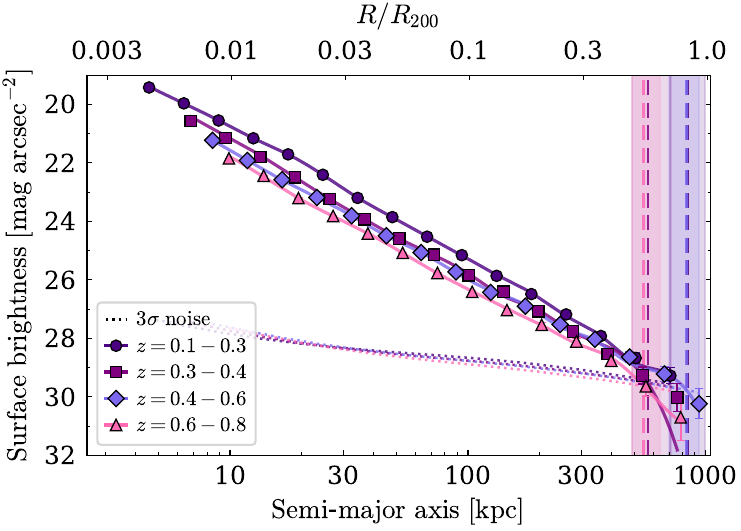}
    \hfill 
    \includegraphics[width=0.5\linewidth]{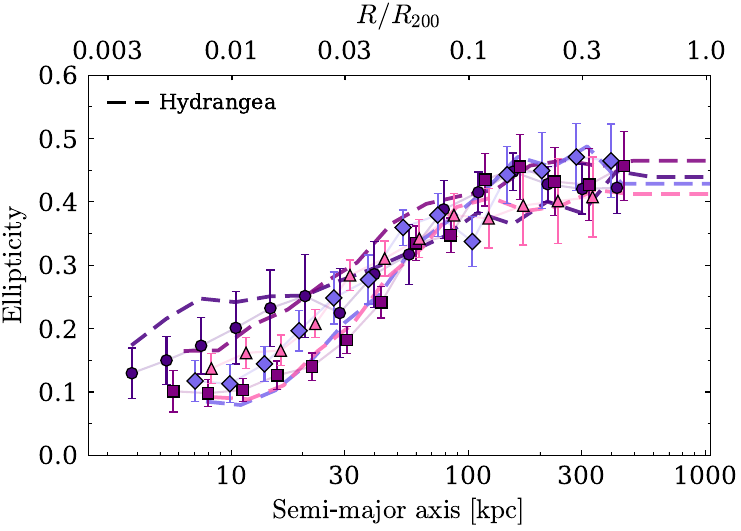}
    \caption{Surface brightness and ellipticity profiles derived from stacked \HE-band images along the semi-major axis of clusters in different redshift bins. Prior to stacking, the member galaxies and bright sources were masked, and the position angles of the clusters were aligned to the $x$ axis as explained in Sects.~\ref{sec:methods:Data_extraction_BCG_ICL_shapes} and \ref{sec:average_shape_profiles_stacked_images}. To ensure consistency across redshift bins, all radii were converted to physical units (kiloparsecs) using the midpoint of each redshift bin for angular scale conversion. The clusters used in these distributions span a halo mass range of $1\leq M_{200\mathrm{c}}/10^{14}\,M_{\odot} \leq 3$. \emph{Left}: Surface brightness measured in a series of elliptical annuli (coloured markers representing redshifts) and their linear interpolation (solid lines). Dotted lines indicate the interpolated noise profiles for each bin (see Sect.~\ref{sec:methods:Data_extraction_BCG_ICL_shapes}) . Vertical dashed lines with shaded regions mark $R_{3\sigma}$ and its uncertainties (16th and 84th percentiles). \emph{Right}: Corresponding ellipticity profiles. Dashed lines show the ellipticity distributions obtained from stacked Hydrangea cluster images at redshifts of 0.2, 0.4, 0.5, and 0.7; colours match the closest redshift bin in observations.}
    \label{fig:Ellip_SB_profile_stacks}
\end{figure*}

In the previous sections we have demonstrated that we can measure the ellipticity, PA, and surface brightness profiles of the ICL in individual clusters out to $\sim$\,$0.1\,R_{200\mathrm{c}}$ in Euclid Q1 images. To test whether additional insight can be gained from probing the ICL to larger radii, we performed a stacking analysis. Each cluster image is given equal weight in the stack, but to minimise potential biases, we restricted our analysis to clusters within a mass range of $1 \leq M_{200\mathrm{c}}/10^{14}\,M_\odot \leq 3$, yielding a subsample of 163 clusters. This subsample is further divided into four redshift bins: 0.1--0.3 (31 clusters), 0.3--0.4 (39 clusters), 0.4--0.6 (41 clusters), and 0.6--0.8 (52 clusters).

We constructed a stacked image of the BCG+ICL emission from the clusters in each redshift bin by stacking the background-subtracted and masked individual cluster images. To preserve the shape information as best as possible, the individual cluster images were aligned according to the PA of their ICL measured at $0.1\,R_{200\mathrm{c}}$. We adopted this approach because our primary interest is the ellipticity and orientation of the outer ICL component, rather than the inner regions dominated by the BCG, whose PA can differ from that of the large-scale diffuse light distribution (Fig.~\ref{fig:Average_ellipticity_PA}). For 75 clusters, we are unable to measure the PA of the ICL at $0.1\,R_{200\mathrm{c}}$, so we took the PA at the greatest distance at which it can be measured. The major axis of the ICL in the stacked images is therefore, by construction, aligned with the $x$ axis. We then computed the mean of each pixel in the aligned images, using $3\sigma$ clipping to remove outliers from, for example, imperfect masking. 

Figure~\ref{fig:H_stacked_images} presents the stacked \HE-band images for each redshift bin, with white solid and cyan dashed circles drawn at $R_{200\mathrm{c}}$ and $0.4\,R_{200\mathrm{c}}$, respectively, where $R_{200\mathrm{c}}$ is defined as the mean $R_{200\mathrm{c}}$ values of clusters within each redshift bin. Magenta contours trace the isophotes of stacked flux distribution, smoothed with a Gaussian kernel of $\sigma=30$ pixels ($\approx 9''$), providing a visual guide to the projected spatial extent of the ICL.

The surface brightness and ellipticity profiles of the stacked images are shown in Fig.~\ref{fig:Ellip_SB_profile_stacks}. 
The elliptical isophotes of BCG+ICL light were extracted from the stacked images with \texttt{AutoProf}. The noise in the surface brightness profile is taken as that calculated in the EDF-S field with a background box size corresponding to 1\,Mpc calculated with an angular diameter distance from the midpoint redshift of the bin. This noise profile is then rescaled by the square root of the number of clusters in the stack to reflect the reduction in noise from averaging many clusters. Uncertainties in the ellipticities are derived from bootstrapping (with replacement) the clusters in each stack and measuring the 1\,$\sigma$ scatter in the measured ellipticities of each annulus. 

To interpret the stacked ellipticity profiles, we repeated the same stacking procedure using the forward-modelled Hydrangea mock clusters. We selected mocks in the same halo-mass range as the observed stack ($1 \leq M_{200\mathrm{c}}/10^{14}\,M_\odot \leq 3$), yielding 105 clusters at $z=0.2$, 117 at $z=0.4$, 99 at $z=0.5$, and 183 at $z=0.7$. As for the analysis of observed clusters, the masked and background-subtracted mock images were aligned by the PA of the ICL prior to stacking. Ellipticity profiles were extracted from each stacked mock image using \texttt{AutoProf}, which are shown as the dashed lines in the right panel of Fig.~\ref{fig:Ellip_SB_profile_stacks}; each curve is coloured to match the closest observational redshift bin.

As expected, by stacking the clusters we are able to detect the ICL to much larger distances than in the individual clusters. The most extended ICL signal is observed in the redshift bins 0.1--0.3 and 0.4--0.6, where we can measure the ICL signal above the ${3\sigma}$ noise level out to ${\sim}0.8\,R_{200\mathrm{c}}$, corresponding to a physical extent of $\sim$800\,kpc. In the 0.3--0.4 and 0.6--0.8 redshift bins, the ICL signal extends beyond ${\sim}0.55$--$0.6\,R_{200\mathrm{c}}$, corresponding to ${\sim} 600\,\textrm{kpc}$.

Furthermore, as shown in the right panel of Fig.~\ref{fig:Ellip_SB_profile_stacks}, the stacked profiles enable us to trace the ICL ellipticity to much larger radii than in the individual clusters, reaching beyond $0.3\,R_{200\mathrm{c}}$, which corresponds to physical scales of approximately 300--400\,kpc, depending on the redshift bin. 
Across redshift, the ellipticity profiles follow a similar trend: the central regions are rounder, with ellipticities of 0.1 to 0.2 at 10\,kpc, increasing with radius to reach ellipticities of 0.40--0.45 beyond $0.1\,R_{200\mathrm{c}}$. This trend mirrors the ellipticity profile found in the individual clusters (see Fig.~\ref{fig:Average_ellipticity_PA}), but the ellipticities derived from the stacked images are slightly rounder, reaching up to $\sim$0.40--0.45 at $0.1\,R_{200\mathrm{c}}$, rather than $\sim$0.5 for the median ellipticity of the individual profiles. We attribute this difference primarily to small residual misalignments in the measured PAs of individual clusters prior to stacking, which can partially smooth the intrinsic ellipticity signal, as well as to the increased background homogenisation introduced through the stacking process.

Despite expectations from previous studies that dark matter halo ellipticities should evolve with redshift \citep{Ho_2006,Cataldi_2023}, becoming rounder at lower redshifts as clusters dynamically relax, we do not observe any significant redshift dependence in the \Euclid ellipticity profiles. As shown in the right panel of Fig.~\ref{fig:Ellip_SB_profile_stacks}, the observed \Euclid clusters reach a consistent plateau of ellipticity at $R>0.1\,R_{200\mathrm{c}}$ and we observe no correlation between the redshift of the cluster and the ellipticity. Similarly, the stacked profiles of the simulated clusters also show no significant redshift trend, in contrast to theoretical expectations. It is possible that the number of clusters within the Q1 sample in each redshift bin may be insufficient to reveal the predicted evolution, and that the stacking procedure combined with the limited number of clusters per redshift bin potentially dilutes any underlying signal. We therefore leave a more robust analysis of how the ellipticity of the ICL varies with redshift to DR1, when thousands of clusters will be observed.

\section{Alignment of the ICL with satellite galaxy distribution}
\label{sec:shape_alignment_BCG_ICL_galaxies}

Previous studies have reported a close alignment between the PA of the ICL and the surrounding satellite galaxy distribution \citep[e.g.][]{Huang_2016,West_2017,Kluge_2021}. To test whether the \Euclid clusters exhibit similar behaviour, we measured and compared the PA alignment of the ICL with both the BCG and the satellite galaxy distribution. The galaxy member catalogues were taken from the eRASS1\footnote{eRASS1 cluster member data can be accessed at \url{https://erass-cluster-inspector.com/index.php}} and DES\footnote{DES cluster member catalogues can be found at \url{https://des.ncsa.illinois.edu/releases/y3a2/Y3key-cluster}} databases.

To ensure that cluster shapes are measured within dynamically relaxed regions, we restricted the analysis to member galaxies located within a projected radius of 500\,kpc from the BCG. This threshold follows the approach of \citet{West_2017} and focuses on the region most physically associated with the BCG and the ICL. On average, there are 19 member galaxies within 500\,kpc from the BCGs in our full cluster sample.

The distribution of the satellites was quantified using the shape tensor, weighted by the membership probability $P_{\rm mem}$ provided in the cluster catalogues. We note that both the eRASS and DES catalogues use redMaPPer to define $P_{\rm mem}$, ensuring a consistent weighting scheme across our full cluster sample.\footnote{Definition of $P_{\rm mem}$ within the redMaPPer algorithm is described in \citet{Rykoff_2014}.} Following \citet{zemp_determining_2011}, the shape tensor was defined as
\begin{equation}\label{eqn:tensor}
S_{ij} = \frac{\sum_n P_{{\rm{mem}}, n} x_{n,i} x_{n,j}}{\sum_n P_{{\rm{mem}}, n}},
\end{equation}
where $i$ and $j$ index the two projected spatial axes, $x_{n,i}$ is the $i{\rm{th}}$ coordinate of the $n{\rm{th}}$ galaxy relative to the BCG position. Diagonalising $S_{ij}$ yields eigenvectors that give the principal axis orientation and eigenvalues proportional to the squared axis lengths $a$ and $b$, where $a>b$. From these, the ellipticity is computed by $e = 1 - b/a$. The tensor was initially applied to the galaxies within a circular aperture of 500\,kpc and the parameters returned were used to define a new elliptical aperture. Further, $S_{ij}$ was recomputed for the members within this aperture, and the process was repeated until the difference in successive axis ratios was less than 0.5\%. This returned a single ellipticity and PA for each cluster.

Uncertainties on ellipticity and PA were estimated using a bootstrap resampling approach. For each cluster, member galaxies were resampled with replacement $10^{5}$ times and the corresponding shape parameters were recalculated for each realisation. The median values of the bootstrap distributions were adopted as the best estimates, while the standard deviations were taken as the $1\sigma$ uncertainties, providing a robust measure of the statistical variability arising from the finite sampling of cluster members. For further analysis in studying ICL-galaxy alignment, we retained only clusters with PA uncertainties smaller than $30^{\circ}$. This criterion leads to a sample of 57 clusters, corresponding to $\sim$\,$30\%$ of our sample, for which the cluster shape can be robustly measured using satellite galaxies. In comparison, the ICL allows for reliable shape measurements in 130 clusters, which are the clusters that have $R_{\rm shape}>0.1\,R_{200\mathrm{c}}$, corresponding to $\sim$\,63\% of the sample.

\begin{figure}
\centering
\includegraphics[width=1\linewidth]{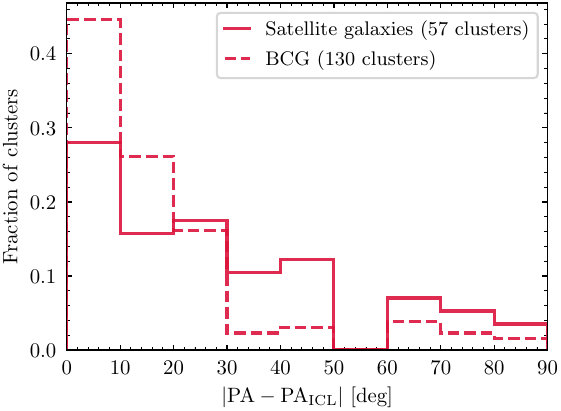}
\caption{Alignment of the ICL with the BCG and the satellite galaxy distribution. The solid line represents the PA difference between the ICL and galaxies, while the dashed line represents the difference between the ICL and the BCG.}
\label{fig:PA_difference_bcg_icl_galaxies}
\end{figure}

Figure~\ref{fig:PA_difference_bcg_icl_galaxies} shows the distributions of PA differences between the ICL and the satellite galaxy distribution (solid line) for the 57 clusters with robust PA measurements of the satellite galaxy distribution, and between the ICL and the BCG (dashed line) for 130 galaxy clusters. For this subsample, 83\% of the satellite galaxies have a membership probability \(P_{\rm mem} > 0.5\), and 68\% have \(P_{\rm mem} > 0.8\). This indicates that the vast majority of galaxies used to define the shape of the satellite galaxy distribution are highly likely to be genuine cluster members, with only a small fraction being potential background contaminants. As a result, contamination from non-member galaxies is unlikely to significantly circularise the shape or bias the PA-difference distributions.

Clusters with $R_{\rm shape}<0.1\,R_{200\mathrm{c}}$ are excluded from these distributions. To account for the inherent $180^{\circ}$ symmetry in PAs, the differences are folded into the range [0$^{\circ}$, 90$^{\circ}$]. The ICL-BCG distribution is strongly skewed towards small PA offsets, with a median misalignment of only $\sim$\,11$^{\circ}$. Approximately 45\% of clusters show ICL-BCG alignment within $10^{\circ}$, and 87\% lie within $30^{\circ}$. These results are broadly consistent with \citet{Gonzalez_2005}, who reported $\sim$\,40\% of their systems aligned within $10^{\circ}$. In contrast, the ICL-galaxy alignments are weaker, with a median offset of $22^{\circ}$; 28\% of clusters are aligned within $10^{\circ}$ and 61\% within $30^{\circ}$. KS tests confirm that both distributions significantly deviate from uniformity, indicating strong ICL-BCG alignment and relatively weaker yet non-random ICL-galaxy alignment. 

Our results are consistent with the ICL-galaxy alignment strengths reported by \citet{West_2017}, who examined 52 clusters up to $z\sim1.8$ and found that $\sim$\,62\% of their sample exhibited $\Delta {\rm PA}$\,$<$\,30$^{\circ}$ when using galaxies within a projected distance of 500\,kpc from the BCG (and by applying a similar PA uncertainty threshold). In contrast, \citet{Kluge_2021} reported stronger alignments, with $\sim$\,80\% of clusters showing ICL-galaxy offsets below $30^{\circ}$ based on spectroscopically confirmed (or high-probability photometric) members smoothed out to $\sim$\,2\,Mpc and with inner position angles matched to the ICL measurement radii. The smaller physical scale adopted in our study isolates the dynamically relaxed cluster core, similar to the approach of \citet{West_2017}, whereas the larger aperture used by \citet{Kluge_2021} traces more extended substructures and filaments that can distort the measured galaxy orientations.

We find that omitting the PA uncertainty threshold applied to isolate clusters with reliably measured galaxy shapes results in a much flatter ICL-galaxy alignment distribution across the sample. This indicates that the weakening of the alignment we observe is primarily driven by poorly constrained satellite galaxy shape measurements rather than by a true physical misalignment. In clusters with a few confirmed members or ongoing dynamical activity, the satellite distribution becomes irregular and the derived PAs highly uncertain, masking any underlying correlation with the ICL. This behaviour reinforces the view that the ICL provides a more stable and coherent tracer of the underlying dark matter potential than the satellite galaxies. At higher redshifts, where cluster membership identification relies more heavily on photometric redshifts and as the spectroscopic completeness decreases, the diffuse ICL -- being less sensitive to small-number statistics and substructure -- is expected to trace the gravitational potential more reliably than the satellite galaxy distribution.

\section{Conclusions}
\label{sec:conclusion}
We have characterised the shape of the ICL haloes for 205 galaxy clusters using Q1 images in five photometric bands (\IE, \YE, \JE, \HE, and $YJ\HE$). We summarise our findings as follows:
\begin{enumerate}
    \item The shape of the ICL halo for clusters at $z>0.2$ is best measured in the \HE or combined $YJ\HE$ bands of \Euclid images, while for clusters at $z<0.2$, the \IE band may be better. Though the \Euclid \IE\ images are deeper, the flux of the ICL peaks in the NIR wavelengths and is characterised by a red colour ($\IE-\HE\sim 1.5$; \citealt{Q1-SP008}), so the \HE images enabled us to measure the ICL profile out to larger radii, especially at higher redshifts. The ICL ellipticity measurements are consistent across all the measured bands: Ellipticity differences relative to the \HE band remain within $\pm0.3$ in 90\% of clusters.
    
    \item We were able to measure the ICL halo PA and ellipticities at $0.1\,R_{200\mathrm{c}}$ ($\sim$\,100\,kpc for the typical cluster in our sample) in individual \HE images for 63\% of our sample. The ICL haloes have a median ellipticity of $\sim$\,0.5. The sample exhibits a broad range of ellipticities such that the distribution has a scatter of $\sigma$\,$\simeq$\,0.18. This distribution is in close agreement with halo ellipticities inferred from both strong and weak lensing studies. The consistency between the ICL and lensing results suggests that ICL measurements at $0.1\,R_{200\mathrm{c}}$ effectively probe the cluster halo shape.
    
    \item  The ellipticity of the BCG+ICL haloes varies with radius. On average, the BCG in the centre ($0.01\,R_{200\mathrm{c}}$) is round, with ellipticities of $\sim$\,0.15--0.20. The surrounding halo then becomes gradually more elliptical with radius, reaching $\sim$\,0.5 at $0.1\,R_{200\mathrm{c}}$.
    
    \item Stacking clusters (aligned by the ICL PA) pushes the ICL detection to ${\sim} 0.6$--$0.8\,R_{200\mathrm{c}}$ and lets us trace ellipticity robustly to ${\sim}0.3$--$0.4\,R_{200\mathrm{c}}$. The stacked ellipticity is slightly rounder than that of individual clusters ($\sim 0.4$--$0.45$ beyond $0.1\,R_{200\mathrm{c}}$ versus ${\sim} 0.5$) because residual PA misalignments and isophotal twists bias stacks low. We find that the ellipticity of the ICL plateaus after $0.1\,R_{200\mathrm{c}}$, and contrary to theoretical expectations, we observed no correlation between ICL ellipticity and redshift over $0.1\le z \le 0.8$. 
      
    \item The ICL and BCG exhibit a strong alignment, with 87\% of clusters aligned within $30^{\circ}$, whereas the ICL-satellite galaxy alignment is slightly weaker, with only 61\% of clusters aligned within $30^{\circ}$. However, robust halo shapes can only be measured via satellite distributions for $\sim$30\% of our sample, whereas ICL measurements reveal halo shapes for 63\% of the sample. This indicates that the ICL PA is both easier to measure and, consequently, a more reliable tracer of halo shape than photometrically selected satellites, most notably at high redshift.
    
\end{enumerate}

Our study demonstrates the capability of \Euclid observations to systematically characterise the morphology of ICL in galaxy clusters, and establish it as a proxy for halo shape. Looking ahead, ICL-derived halo shapes will provide a key observable for future science enabled by upcoming \Euclid data releases. They enable the selection of dynamically relaxed systems, provide robust PAs for stacking and deprojection, reduce scatter in mass--observable relations by constraining 3D orientations, and trace recent accretion geometry to interpret BCG growth and environmental quenching. With the depth and homogeneity of \Euclid, shape- and PA-based cuts can be applied uniformly across redshift to optimise cluster samples and tighten mass calibration, putting ICL morphology to work for survey cosmology and galaxy-evolution research.

\vspace{1cm} 

\begin{acknowledgements}
T. Kolcu, N. A. Hatch, A. Fernández, H. Gao, J. B. Golden-Marx, H. Gully, M. Butler and M. Urbano gratefully acknowledge financial support from The Leverhulme Trust. 

MM acknowledges support from grant RYC2022-036949-I financed by the MICIU/AEI/10.13039/501100011033 and by ESF+, grant PID2024-158845NB-I00 financed by MICIU/AEI/10.13039/501100011033 and by ERDF, EU, and program Unidad de Excelencia Mar\'{i}a de Maeztu CEX2020-001058-M.

YMB acknowledges support from UK Research and Innovation through a Future Leaders Fellowship (grant agreement MR/X035166/1) and financial support from the Swiss National Science Foundation under project 200021\_213076. This work used the DiRAC@Durham facility managed by the Institute for Computational Cosmology on behalf of the STFC DiRAC HPC Facility (www.dirac.ac.uk). The equipment was funded by BEIS capital funding via STFC capital grants ST/K00042X/1, ST/P002293/1, ST/R002371/1 and ST/S002502/1, Durham University and STFC operations grant ST/R000832/1. DiRAC is part of the National e-Infrastructure.

FD acknowledges support from CNES.

AHG acknowledges the support of NASA ROSES Grant 12-EUCLID11-0004.

LM acknowledges the financial contribution from the PRIN-MUR 2022 20227RNLY3 grant ``The concordance cosmological model: stress-tests with galaxy clusters'' supported by Next Generation EU and from the grant ASI n. 2024-10-HH.0 ``Attività scientifiche per la missione \Euclid - fase E''.

RR acknowledges financial support through grants PRIN-MIUR 2020SKSTHZ and through INAF WEAVE StePS funds.

AS is supported by ICSC, project funded by European Union - NextGenerationEU - and National Recovery and Resilience Plan (NRRP) - Mission 4 Component 2, the PRIN 2022 PNRR project (202259YAF), PRIN 2022 (20225E4SY5), INAF Large Grant 2023 “Witnessing the Birth of the Most Massive Structures of the Universe”, and the INFN Indark Grant.

\AckERO \AckQone
\AckDatalabs
\AckEC \AckCosmoHub

\end{acknowledgements}

\bibliography{references}

\begin{appendix}
\end{appendix}

\label{LastPage}

\end{document}